\documentclass[twocolumn]{aastex631}

\let\tablenum\relax
\usepackage{siunitx}
\usepackage{multirow}
\usepackage{amsmath}
\usepackage{graphicx}

\DeclareRobustCommand{\ion}[2]{%
\relax\ifmmode
\ifx\testbx\f@series
{\mathbf{#1\,\mathsc{#2}}}\else
{\mathrm{#1\,\mathsc{#2}}}\fi
\else\textup{#1\,{\mdseries\textsc{#2}}}%
\fi}

\shorttitle{Decoding Galaxy SEDs}
\shortauthors{Gao et al.}

\begin{document}
\title{From Halos to Galaxies. X: Decoding Galaxy SEDs with Physical Priors and Accurate Star Formation History Reconstruction}

\author[0000-0002-0182-1973]{Zeyu Gao}
\affiliation{Department of Astronomy, School of Physics, Peking University, 5 Yiheyuan Road, Beijing 100871, China}
\affiliation{Kavli Institute for Astronomy and Astrophysics, Peking University, Beijing 100871, China}

\author{Yingjie Peng}
\affiliation{Department of Astronomy, School of Physics, Peking University, 5 Yiheyuan Road, Beijing 100871, China}
\affiliation{Kavli Institute for Astronomy and Astrophysics, Peking University, Beijing 100871, China}

\author[0000-0002-3775-0484]{Kai Wang}
\affiliation{Kavli Institute for Astronomy and Astrophysics, Peking University, Beijing 100871, China}
\affiliation{Institute for Computational Cosmology, Department of Physics, Durham University, South Road, Durham, DH1 3LE, UK}
\affiliation{Centre for Extragalactic Astronomy, Department of Physics, Durham University, South Road, Durham DH1 3LE, UK}

\author[0000-0001-6947-5846]{Luis C. Ho}
\affiliation{Kavli Institute for Astronomy and Astrophysics, Peking University, Beijing 100871, China}
\affiliation{Department of Astronomy, School of Physics, Peking University, 5 Yiheyuan Road, Beijing 100871, China}

\author[0000-0002-7093-7355]{Alvio Renzini}
\affiliation{INAF--Osservatorio Astronomico di Padova, Vicolo dell’Osservatorio 5, I-35122 Padova, Italy}

\author[0000-0002-9656-1800]{Anna R. Gallazzi}
\affiliation{INAF--Osservatorio Astrofisico di Arcetri, Largo Enrico Fermi 5, I-50125 Firenze, Italy}

\author[0000-0002-4803-2381]{Filippo Mannucci}
\affiliation{INAF--Osservatorio Astrofisico di Arcetri, Largo Enrico Fermi 5, I-50125 Firenze, Italy}

\author[0000-0001-5356-2419]{Houjun Mo}
\affiliation{Department of Astronomy, University of Massachusetts, Amherst MA 01003, USA}

\author[0000-0002-4534-3125]{Yipeng Jing}
\affiliation{Department of Astronomy, School of Physics and Astronomy, Shanghai Jiao Tong University, Shanghai 200240, China}
\affiliation{Tsung-Dao Lee Institute, and Shanghai Key Laboratory for Particle Physics and Cosmology, Shanghai Jiao Tong University, Shanghai 200240, China}

\author[0000-0003-3997-4606]{Xiaohu Yang}
\affiliation{Tsung-Dao Lee Institute, and Shanghai Key Laboratory for Particle Physics and Cosmology, Shanghai Jiao Tong University, Shanghai 200240, China}
\affiliation{Department of Astronomy, School of Physics and Astronomy, Shanghai Jiao Tong University, Shanghai 200240, China}

\author[0000-0003-1588-9394]{Enci Wang}
\affiliation{CAS Key Laboratory for Research in Galaxies and Cosmology, Department of Astronomy, University of Science and Technology of China, Hefei, Anhui 230026, China}
\affiliation{School of Astronomy and Space Science, University of Science and Technology of China, Hefei 230026, China}

\author[0009-0001-1564-3944]{Dingyi Zhao}
\affiliation{Department of Astronomy, School of Physics, Peking University, 5 Yiheyuan Road, Beijing 100871, China}
\affiliation{Kavli Institute for Astronomy and Astrophysics, Peking University, Beijing 100871, China}

\author[0000-0002-6961-6378]{Jing Dou}
\affiliation{School of Astronomy and Space Science, Nanjing University, Nanjing 210093, People’s Republic of China}

\author[0000-0002-3890-3729]{Qiusheng Gu}
\affiliation{School of Astronomy and Space Science, Nanjing University, Nanjing 210093, People’s Republic of China}

\author[0009-0000-7307-6362]{Cheqiu Lyu}
\affiliation{Department of Astronomy, School of Physics, Peking University, 5 Yiheyuan Road, Beijing 100871, China}
\affiliation{Kavli Institute for Astronomy and Astrophysics, Peking University, Beijing 100871, China}
\affiliation{CAS Key Laboratory for Research in Galaxies and Cosmology, Department of Astronomy, University of Science and Technology of China, Hefei, Anhui 230026, China}
\affiliation{School of Astronomy and Space Science, University of Science and Technology of China, Hefei 230026, China}

\author[0000-0002-4985-3819]{Roberto Maiolino}
\affiliation{Cavendish Laboratory, University of Cambridge, 19 J.J. Thomson Avenue, Cambridge, CB3 0HE, UK}
\affiliation{Kavli Institute for Cosmology, University of Cambridge, Madingley Road, Cambridge, CB3 0HA, UK}
\affiliation{Department of Physics and Astronomy, University College London, Gower Street, London WC1E 6BT, UK}

\author[0000-0002-6137-6007]{Bitao Wang}
\affiliation{School of Physics and Electronics, Hunan University, Changsha 410082, China}
\affiliation{Kavli Institute for Astronomy and Astrophysics, Peking University, Beijing 100871, China}

\author[0000-0002-8429-7088]{Yu-Chen Wang}
\affiliation{Department of Astronomy, School of Physics, Peking University, 5 Yiheyuan Road, Beijing 100871, China}
\affiliation{Kavli Institute for Astronomy and Astrophysics, Peking University, Beijing 100871, China}

\author{Bingxiao Xu}
\affiliation{Kavli Institute for Astronomy and Astrophysics, Peking University, Beijing 100871, China}

\author[0000-0003-3564-6437]{Feng Yuan}
\affiliation{Center for Astronomy and Astrophysics and Department of Physics, Fudan University, 2005 Songhu Road, Shanghai 200438, China}

\author[0000-0002-9529-1044]{Xingye Zhu}
\affiliation{Department of Astronomy, School of Physics, Peking University, 5 Yiheyuan Road, Beijing 100871, China}
\affiliation{Kavli Institute for Astronomy and Astrophysics, Peking University, Beijing 100871, China}

\correspondingauthor{Yingjie Peng}
\email{yjpeng@pku.edu.cn}

\begin{abstract}
The spectral energy distribution (SED) of galaxies is essential for deriving fundamental properties like stellar mass and star formation history (SFH). However, conventional methods, including both parametric and nonparametric approaches, often fail to accurately recover the observed cosmic star formation rate (SFR) density due to oversimplified or unrealistic assumptions about SFH and their inability to account for the complex SFH variations across different galaxy populations. To address this issue, we introduce a novel approach that improves galaxy broadband SED analysis by incorporating physical priors derived from hydrodynamical simulations. Tests using IllustrisTNG simulations demonstrate that our method can reliably determine galaxy physical properties from broadband photometry, including stellar mass within \SI{0.05}{dex}, current SFR within \SI{0.3}{dex}, and fractional stellar formation time within\SI{0.2}{dex}, with a negligible fraction of catastrophic failures. When applied to the Sloan Digital
Sky Survey (SDSS) main photometric galaxy sample with spectroscopic redshift, our estimates of stellar mass and SFR are consistent with the widely used MPA-JHU and GSWLC catalogs. Notably, using the derived SFHs of individual SDSS galaxies, we estimate the cosmic SFR density and stellar mass density with remarkable consistency to direct observations up to $z \sim 6$.  This demonstrates a significant advancement in deriving SFHs from SEDs that closely align with observational data. Consequently, our method can reliably recover observed spectral indices such as  $D_{\rm n}(4000)$ and $\rm H\delta_{\rm A}$ by synthesizing the full spectra of galaxies using the estimated SFHs and metal enrichment histories, relying solely on broadband photometry as input. Furthermore, this method is extremely computationally efficient compared to conventional approaches.
\end{abstract}

\keywords{Galaxy evolution (594) --- Galaxy properties (615) --- Galaxy stellar content (621) --- Galaxy colors (586) --- Galaxy ages (576) --- Galaxy masses (607) --- Star formation (1569) --- Galaxy quenching (2040)}

\section{Introduction}%
\label{sec:introduction}

The spectral energy distribution (SED) of galaxies encodes valuable information on a number of astrophysical processes of galaxy evolution, including gas inflow, star formation, gas outflow, and chemical enrichment. Extracting this information effectively from galaxy SEDs is a central challenge in understanding galaxy formation and evolution \citep[see][for a review]{Conroy_2013}. Traditional approaches typically involve forward-modeling the synthesized SED and adjusting model parameters to fit the observed data of a galaxy.

Deriving accurate star formation history (SFH) is one of the holy grails in SED fitting. Currently, the most widely used tools for obtaining SFH through SED fitting are Bagpipes \citep{Carnall_2018, Carnall_2019a} and Prospector \citep{Leja_2017, Johnson_2021}.
Despite its utility, the SED alone often fails to tightly constrain SFHs due to degeneracies among the effects of different stellar ages, metallicities, and dust extinction \citep{Faber_1972, O'Connell_1976, Worthey_1994, Carnall_2019a, Papovich_2001}, and the outshining problem \citep[i.e., young stars outshine their older counterparts, making it hard to constrain the old stellar populations from SEDs;][]{Papovich_2001}. Observational limitations such as restricted wavelength coverage and low signal-to-noise ratios further complicate this issue.

Strong priors have been implemented on the SFHs to mitigate these challenges. Common approaches include parametric models like the exponential model
$(\propto e^{-(t - t_0)/\tau})$, which is also known as the $\tau$ model and was widely used, perhaps for being the result of the closed-box model in which
the star formation rate (SFR) is proportional to the gas mass \citep{Schmidt_1959}, which is assumed to be all in place at the
beginning. This is not what happens to real galaxies, which start from small seeds and grow secularly by merging and gas accretion. Moreover, a single function is incapable of capturing features from recent
starbursts and, to cope with this, additional burst components were incorporated \citep{Kauffmann_2003, Lee_2009}. In an attempt to use more realistic SFHs, exponentially rising 
($\propto e^{(t - t_0) / \tau}$) and delayed-exponential ($\propto (t - t_0)e^{-(t - t_0) / \tau}$) relations were introduced 
\citep{Maraston_2010, Lee_2010}. 
Motivated by the coverage of observational parameter space of spatially resolved spectra from the Calar Alto Legacy Integral Field Area (CALIFA) survey \citep{Sanchez_2012}, a library of SFHs modeled as double-Gaussian plus random bursts was also applied \citep[see][as well as a high redshift application in preparation]{Zibetti_2017, Zibetti_2020}. 
Motivated by the shape of the cosmic SFR density \citep{Madau_2014}, the $\Gamma$ function was also used in the SED fits \citep{Lu_2015, Zhou_2020}; with others preferring the log-normal function or a double power law \citep[see][]{Gladders_2013, Abramson_2016, Diemer_2017, Carnall_2018}. Finally, to increase flexibility, the combination of different functions is sometimes also applied \citep[e.g.,][]{Carnall_2018, Han_2023}.

Clearly, there is no guarantee that galaxies have evolved following any of these simple analytical
functions, which introduce arbitrary nonphysical priors \citep{Simha_2014, Carnall_2019a}. Consequently, models based on the nonparametric SFHs have also been
employed \citep{CidFernandes_2005, Ocvirk_2006, Leja_2017, Iyer_2017, Chauke_2018, Iyer_2019}.
Regarding the limited ability to constrain the detailed shape of star formation
histories, these nonparametric methods adopted broad step functions with
additional continuity requirements \citep{Cappellari_2012,
Walcher_2015, Leja_2017, Leja_2019b}. Although these nonparametric methods can
successfully recover simple, artificially input SFHs
\citep{Leja_2019b}, realistic SFHs of actual galaxies
are still challenging. Moreover, nonparametric methods suffer from various degeneracies, are computationally demanding,
and require high-quality data to constrain SFHs.

A promising solution to these limitations involves leveraging physical priors from realistic galaxy formation models. These models, including empirical models
\citep[e.g.,][]{Conroy_2009b, Moster_2010, Yang_2012, Behroozi_2019, Chen_2021},
semi-analytic models \citep[e.g.,][]{White_1991, Kauffmann_1993, Somerville_1999, Guo_2011}, and
hydrodynamical simulations \citep[e.g.,][]{Katz_1992, Crain_2009, Pillepich_2018a, Schaye_2015, Schaye_2023},
should be realistic in the sense that they are capable of reproducing various
distribution functions and scaling relations for observed galaxies. Using these models' star formation and metal enrichment histories, it is possible to establish more physically grounded priors for inferring real galaxy histories from SEDs. Similar approaches have been put into practice in
\citet{Pacifici_2012}, where they use a semi-analytic
galaxy formation model as a prior to infer the physical properties from galaxy
observables \citep[see also][]{Finlator_2007, Pacifici_2016}. Recently, \citet{Zhou_2022} incorporate
the semi-analytic model of galaxy evolution processes, including inflow,
outflow, star formation, and chemical enrichment \citep{Talbot_1971, Tinsley_1974, Chiosi_1980, Tinsley_1980, Lacey_1985, Bouche_2010, Lilly_2013, Dekel_2013, Peng_2014, Dekel_2014, Dou_2021, Wang_2021, Wang_2022},
into the modeling of galaxy spectra.

This work utilizes the SFHs and metal enrichment histories of realistic galaxies in state-of-the-art hydrodynamical simulations
to infer the physical properties of observed galaxies from their broadband
photometry. We demonstrate that our method accurately recovers stellar mass, current SFR, and comprehensive SFHs of
the test galaxy sample. When applied to actual galaxies,
our method can deliver realistic SFHs and metal enrichment histories that can not only reproduce the general trends including cosmic SFR density and cosmic stellar mass density, but
also recover several observed spectral indices, despite relying solely on broadband photometry.

The paper is structured as follows. The data and method are introduced
in \S\,\ref{sec:data} and \S\,\ref{sec:method}, respectively. The test and
validation of our method are presented in \S\,\ref{sec:testing_on_tng100}. Then, we apply it to the observed galaxies and show the results in
\S\,\ref{sec:applications_on_sdss}. Finally, we present the summary in
\S\,\ref{sec:conslusions}. Throughout this paper, we adopt the initial mass function from \citet{Chabrier_2003} and the cosmology
from \citet{PlanckCollaboration_2016}, with $\Omega_{m, 0} = 0.3075$,
$\Omega_{\Lambda, 0}=0.6910$, and $h=0.6774$.

\section{Data}%
\label{sec:data}

\subsection{Simulation}%
\label{sub:simulation}

The IllustrisTNG project encompasses a series of cosmological hydrodynamical
simulations
\citep{Marinacci_2018,Naiman_2018,Nelson_2018,Nelson_2019,Pillepich_2018a,
Pillepich_2018b, Springel_2018}, simulating the evolution of galaxies from
$z \sim 127$ to $z=0$ with the moving-mesh code \textsc{arepo}
\citep{Springel_2010} across three distinct volumes: $35\,h^{-1}\,\mathrm{Mpc}$ for
TNG50, $75\,h^{-1}\,\mathrm{Mpc}$ for TNG100, and $205\,h^{-1}\,\mathrm{Mpc}$
for TNG300. Subgrid recipes are adjusted so that the
stellar mass function, the stellar mass--black hole mass relation, the 
mass--size relation, the cosmic SFR density, the intragroup medium, the
mass--metallicity relation, and the galaxy quenching match the observational
results under the resolution of TNG100. Therefore, we employ TNG100 in our study, which offers a mass resolution of $1.4\times
10^6\, h^{-1}\, M_{\odot}$ for dark matter particles and $2.6\times
10^5\, h^{-1}\, M_{\odot}$ for gas cells. This study focuses on galaxies above
$10^9\, h^{-1}\, M_{\odot}$, so that each galaxy is finely sampled with more than 3800
stellar particles.

Dark matter halos within the simulation are identified using the
friends-of-friends (FoF) algorithm \citep{Davis_1985}. In each FoF halo,
substructures are identified using the \textsc{subfind} algorithm
\citep{Springel_2001, Dolag_2009} using dark matter particles with gas and star particles attached. Each baryonic substructure is designated as a galaxy, and its dark matter counterpart as a subhalo. The subhalo with the minimal gravitational potential is defined as the
central subhalo, and the corresponding galaxy is defined as the central galaxy. Others are classified as satellites. The merger histories are tracked using the \textsc{sublink}
algorithm \citep{Rodriguez-Gomez_2015}, and the main progenitor of each subhalo is defined as the most massive one among all its progenitors in the branching point. \citep{DeLucia_2007}.

Mock Sloan Digital Sky Survey (SDSS) photometry for each galaxy is synthesized from the stellar particles' mass, metallicity, and age, with dust attenuation modeled using the distribution of metal-enriched gas \citep[see][for details]{Nelson_2018}. The stellar population synthesis is performed using the \textsc{fsps} code
\citep{Conroy_2009a, Conroy_2010, Johnson_2022}, incorporating Padova isochrones and the MILES stellar library \citep{Marigo_2007, Marigo_2008, Sanchez-Blazquez_2006}, with an initial mass function from \citet{Chabrier_2003}.

\subsection{Observational data}%
\label{sub:observational_data}

The observational dataset for this study is sourced from the SDSS main galaxy sample of the seventh data release, comprising approximately 210,000 galaxies within the redshift range of $0.02 < z < 0.085$
\citep{York_2000, Abazajian_2009}. The detailed selection criteria can be found in \citet{Peng_2010}.
Each galaxy has five broadband photometric
magnitudes ($u, g, r, i, z$) and a spectroscopic redshift. The rest-frame
magnitudes are obtained by applying the K-correction procedure with the Python
package \textsc{kcorrect} \citep{Blanton_2007}. We also calculate the $V_{\rm
max}$ factor \citep{Schmidt_1968} for each galaxy using the \textsc{kcorrect} package, assessing the maximum redshift at which it can be detected, given the survey's magnitude limits and fiber collision corrections from the NYU-VAGC catalog \footnote{\url{http://sdss.physics.nyu.edu/vagc/}} \citep{Blanton_2005}.

The galaxies are cross-matched with the MPA-JHU\footnote{\url{https://wwwmpa.mpa-garching.mpg.de/SDSS/DR7/}}
\citep{Kauffmann_2003,Brinchmann_2004,Salim_2007} and GSWLC \citep{Salim_2007, Salim_2016, Salim_2018},
using a unique ID tuple (\texttt{MJD}, \texttt{PLATE}, \texttt{FIBERID}). From the MPA-JHU catalog, we obtain two critical spectral indices, $D_{\rm n}(4000)$ and $\rm H\delta_A$, as well as estimates of stellar mass and SFRs. It is important to note that these spectral indices are adjusted for sky-line contamination by fitting the continuum using the model outlined in \citet{Bruzual_2003}. It is also noteworthy that the spectral indices are only used for comparison in \S\S\,\ref{subsec:spec_idx} and \textit{not} an input in the SED-fitting procedure.  

The estimation of stellar mass in the MPA-JHU catalog involves a two-part methodology: the central region covered by the fiber is evaluated using various spectral indices, while the outer region's mass is derived from five broadband magnitudes. SFRs in the MPA-JHU catalog are determined using $\rm H\alpha$ luminosity for star-forming galaxies, and the SFR upper limits are provided by the $D_{\rm n}(4000)$ index for quiescent galaxies and those with active galactic nuclei. Adjustments are made to convert the initial mass function from \citet{Kroupa_2001} to that of \citet{Chabrier_2003}, applying a conversion factor from \citet{Madau_2014} (\SI{-0.034}{dex} for stellar mass and \SI{-0.027}{dex} for SFR). 
In GSWLC, both the
stellar mass and SFR for each galaxy are obtained by
simultaneously fitting the multiband photometry from the IR to UV using the
\text{CIGALE} code \citep{Noll_2009,Boquien_2019} with the synthesis model from \citet{Bruzual_2003} and the initial mass function in
\citet{Chabrier_2003}.

\section{Method}%
\label{sec:method}

We propose a novel method to infer the SFHs, metal enrichment histories, and stellar mass-to-light ratios of galaxies using broadband photometry, with simulated galaxies from hydrodynamical galaxy formation
simulations as a prior. To begin with, we consider the broadband magnitudes
and the corresponding colors for the $i$th simulated galaxy as
\begin{align}
    \tilde {\mathbf m}_i &= (\tilde m_{i, 1}, \tilde m_{i, 2}, ..., \tilde
    m_{i, n})\nonumber\\
    \tilde {\mathbf c}_i &= (\tilde c_{i, 1}, \tilde c_{i, 2}, ..., \tilde
    c_{i, n - 1}),~~~\tilde c_{i, j} \equiv \tilde m_{i, j + 1} - \tilde m_{i,
    j}
\end{align}
and similarly for the $i$th observed galaxy:
\begin{align}
    \mathbf m_i &= (m_{i, 1}, m_{i, 2}, ..., m_{i, n})\nonumber\\
    \mathbf c_i &= (c_{i, 1}, c_{i, 2}, ..., c_{i, n - 1}),~~~c_{i, j} \equiv
    m_{i, j + 1} - m_{i, j}.
\end{align}

We first consider the inference of the SFH of observed galaxies,
as the metal enrichment history and the stellar mass-to-light ratio can be derived
similarly. To proceed, we need to assume that the star formation
histories and metal enrichment histories for real galaxies in our Universe are drawn from the same distributions as those of the simulated galaxies. This assumption is supported by the TNG100 simulation's ability to replicate key observational statistics, including the stellar mass function
\citep[see][]{Pillepich_2018a}, the magnitude--color joint distribution function
\citep[see][]{Nelson_2018}, the cosmic SFR density \citep[see][]{Crain_2023}, the quiescent fraction of galaxies \citep[see][]{Donnari_2021}, and the star-forming main sequence \citep[see][]{Donnari_2019}. However, we emphasize that our method does not require the TNG100 simulation to be strictly correct, as in practice we use it as a set of template SFHs from which to choose the one that best reproduces the observables of individual galaxies. 
The likelihood model is then formulated as:
\begin{equation}
    \mathcal{P}(\mathcal H\mid \mathbf c_i) \propto \mathcal{P}(\mathcal
    H)\times \mathcal{P}(\mathbf c_i\mid \mathcal H)\nonumber =
    \mathcal{P}(\tilde{\mathcal H})\times \mathcal{P}(\mathbf c_i\mid
    \tilde{\mathcal H}),\nonumber
\end{equation}
where $\mathcal H$ and $\tilde{\mathcal H}$ represent the SFHs of observed and simulated galaxies, respectively. Here $\mathbf c_i$
is the multiband color of the $i$th observed galaxy. We estimate the SFH for the $i$th observed galaxy as:
\begin{align}
    \label{eq:sfh_estimate} \hat{\mathcal H}_i(t) &= \frac{\sum_{j=1}^N w_{i,
    j}\tilde{\mathcal H}_j(t)}{\sum_{j=1}^N w_{i, j}}  \\ w_{i, j} &\propto
    \prod_{k=1}^{n-1}\frac{1}{\sigma_{k}}\exp\left(-\frac12 \frac{(c_{i, k} -
    \tilde c_{j, k})^2}{\sigma_{k}^2}\right)\nonumber
\end{align}
where $N$ represents the number of simulated galaxies and $\sigma_{k}$ the uncertainty of the $k$th color. Similar methods can be
employed to estimate the metal enrichment history and the stellar mass-to-light
ratio. A similar approach has been applied to strong emission lines to infer the abundances and electron temperatures in \ion{H}{\textsc{ii}}
 regions \citep[the counterpart method proposed by][]{Pilyugin_2012}.

To assess the reliability of our estimates, we calculate the weighted standard deviation of the prior physical properties as the uncertainty, effectively treating this as the probability density function.

However, as TNG simulations may not perfectly represent the observed galaxies' distribution, minor discrepancies may lead to biases in physical properties for outliers in the color space. To quantify this, we define a weighted distance metric:
\begin{equation}
    \mathcal{D}_i =  \frac{\sum_j w_{i,j}  \sum_{k=1}^{n-1} \left ( c_{i,k} -
    \tilde{c}_{j,k} \right )^2} {\sum_j w_{i,j}} \label{eq:weighted_distance}
\end{equation}
This parameter quantifies the weighted average distance between the $i$th observed
galaxy and the surrounding simulated galaxies in the color space. Galaxies with $\mathcal{D}_i \geq 0.19$ are considered poorly sampled and are excluded from further analysis, affecting approximately 5\% of galaxies in our SDSS DR7 sample. This empirical threshold is chosen because it is the highest $\mathcal{D}_i$ in the self-consistency test; thus, it can somehow be used to describe the limitation of our method.

\section{Testing on the IllustrisTNG simulation}%
\label{sec:testing_on_tng100}

\begin{figure}
    \centering
    \includegraphics[width=0.9\columnwidth]{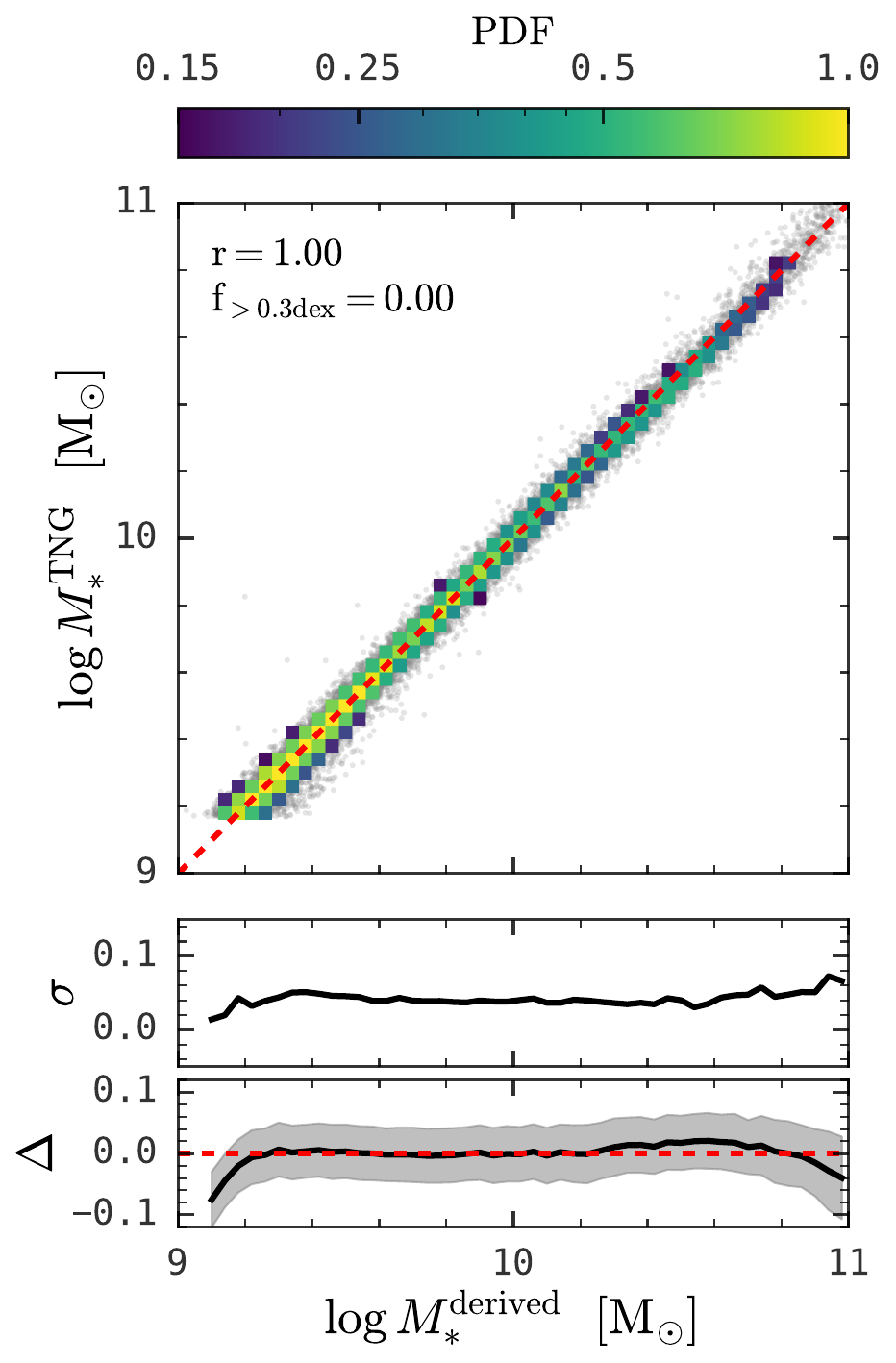}
    \caption{
        Comparison of estimated stellar mass using our method against actual values in TNG100. Individual galaxies are represented by gray dots, and the heat map illustrates the normalized probability distribution. The one-to-one reference line is shown as a red dashed line.
        Spearman's rank correlation coefficient ($r$) and the fraction of
        outliers ($f_{\rm > 0.3 dex}$) are presented in the top-left corner.
        The lower two panels show the standard deviation and the systematic
        bias as a function of estimated stellar mass. The gray-shaded region in the bottom panel shows the median uncertainty. This figure demonstrates
        our method's accuracy ($\approx 0.05$ dex) and negligible systematic bias in estimating stellar masses from broadband optical photometry.
}
    \label{fig:tng_tng_mstar}
\end{figure}

\begin{figure*}
    \centering
    \includegraphics[width=0.9\linewidth]{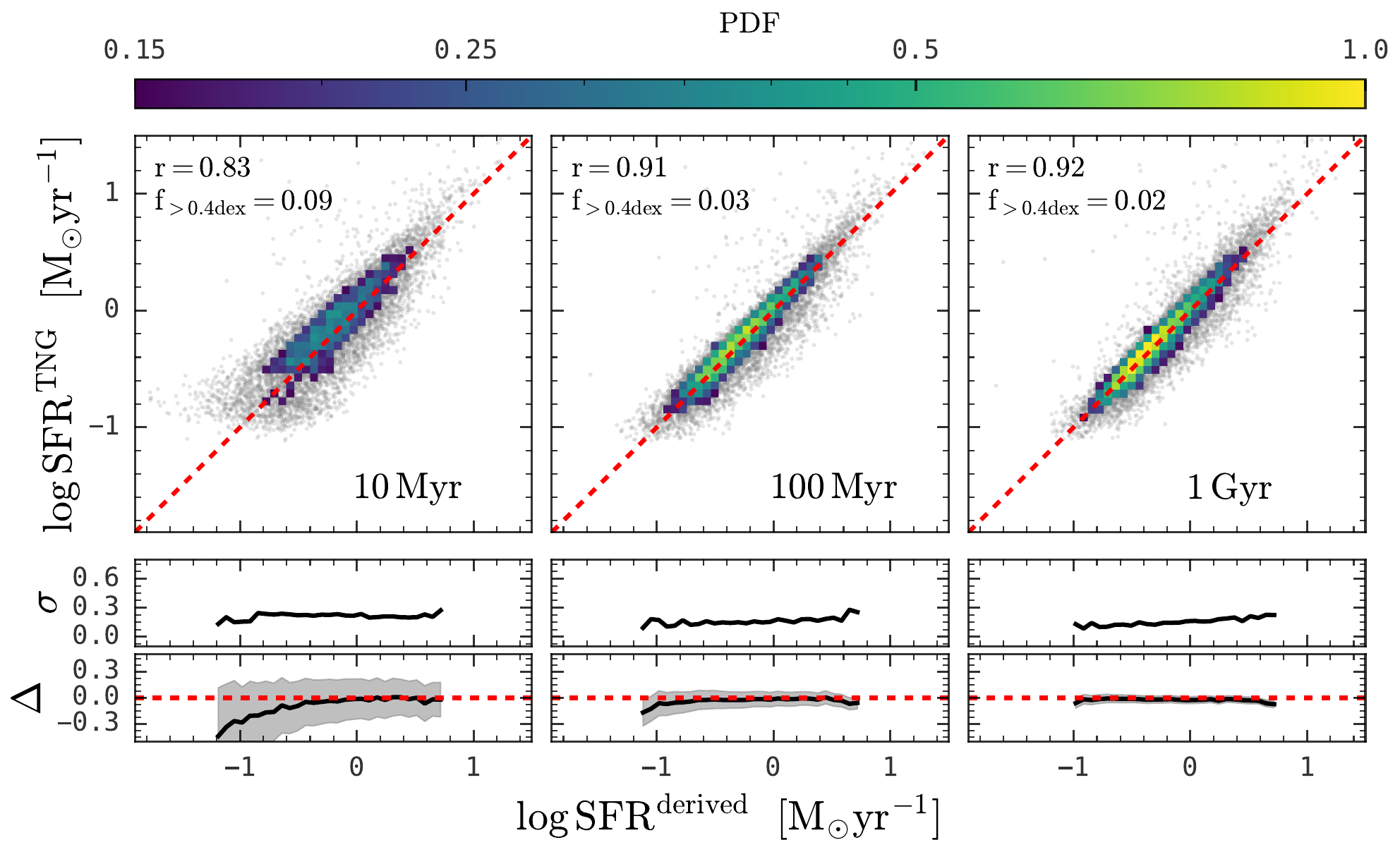}
    \caption{
        Similar to Figure~\ref{fig:tng_tng_mstar}, this figure displays the estimation accuracy of SFRs over three timescales: 10Myr (left), 100Myr (middle), and 1Gyr (right). Gray dots represent individual star-forming galaxies in TNG100, and the heat map illustrates the normalized probability distribution. This
        figure demonstrates our method's effectiveness in recovering SFR with an accuracy of $\lesssim 0.3$ dex and negligible systematic bias.
    }
    \label{fig:tng_tng_sfr}
\end{figure*}

\begin{figure*}
    \centering
    \includegraphics[width=0.9\linewidth]{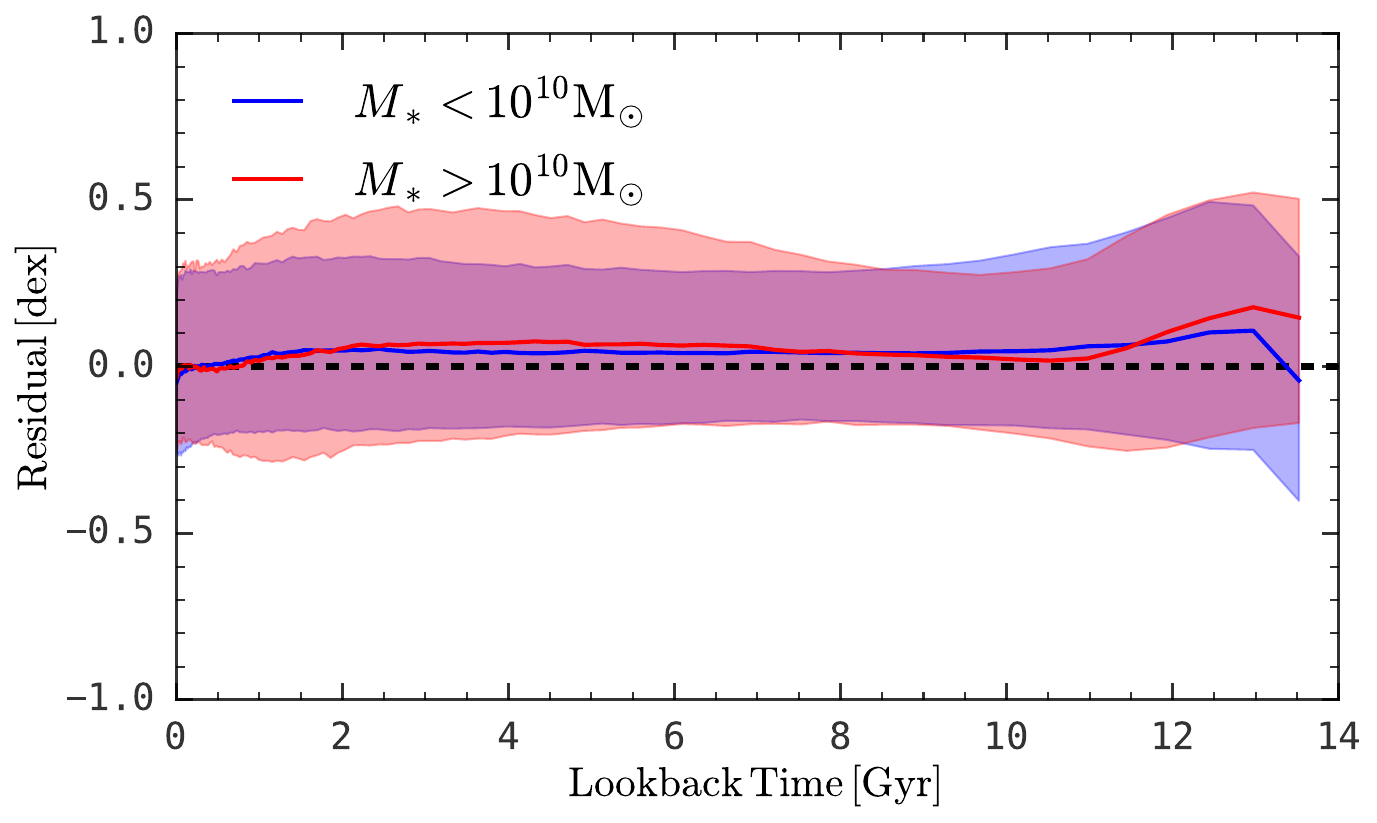}
    \caption{
        The distribution of SFH residuals (derived SFHs minus real SFHs in TNG100). The blue solid line shows the median residual for low-mass galaxies ($M_* > 10^{10} M_{\odot}$), with the blue-shaded region denoting a $1\sigma$ distribution. The high-mass counterparts are shown in red. This figure illustrates our method's capability to estimate galaxy SFHs with about \qty{0.1}{dex} error for the median and \qtyrange{0.3}{0.4}{dex} error when considering the 1-sigma distribution. 
    }%
    \label{fig:tng_tng_sfh}
\end{figure*}

\begin{figure*}
    \centering
    \includegraphics[width=0.9\linewidth]{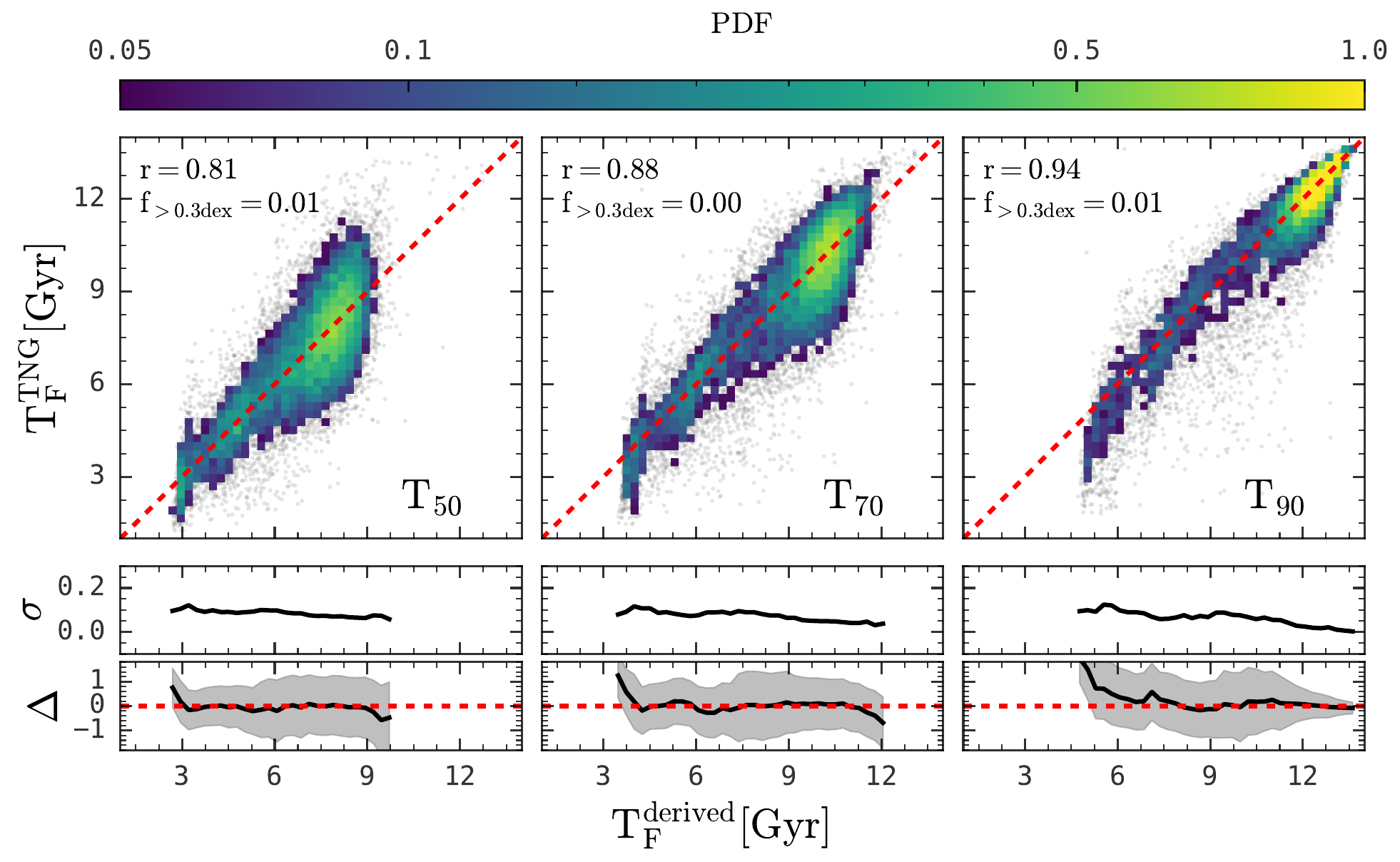}
    \caption{
        Similar to Figure~\ref{fig:tng_tng_mstar}, this figure extends the analysis to the cosmic time when 50\% (left), 70\% (middle), and 90\% (right) of the stellar
        population has formed. Individual
        gray dots represent galaxies in TNG100, and the heat map illustrates the normalized probability distribution. This figure illustrates our method's capability to estimate typical galaxy formation times with a fractional standard deviation of around 30\% and negligible systematic bias.
    }%
    \label{fig:tng_tng_tx}
\end{figure*}

We first test the performance of our method on the simulated galaxies in the
IllustrisTNG simulation before we apply it to the observational galaxy sample.
For each simulated galaxy, it is possible to obtain its SFH, metal enrichment history, and stellar mass-to-light ratio from the stellar particles it
contains. We then apply a self-consistency test with all other simulated galaxies as priors to estimate these properties for the $i$th galaxy.
The accuracy of our method was quantified by comparing these estimates to the properties derived directly from the galaxies' stellar particles.

We first estimate the \textit{r}-band mass-to-light ratio and derive the stellar mass. The results, shown in Figure~\ref{fig:tng_tng_mstar}, indicate that our method reliably estimates stellar masses, with the majority of galaxies exhibiting less than \SI{0.3}{dex} deviation from their true values, and a scatter of less than \SI{0.05}{dex}. We also test another approach based on the \textit{z}-band mass-to-light ratio, which provides similar results.

Similarly, Figure~\ref{fig:tng_tng_sfr} illustrates the method's robustness in accurately estimating the SFRs averaged over three different time scales, which correspond to
different observation probes: \SI{10}{Myr} for indicators based on the $\rm H\alpha$ emission, \SI{100}{Myr} for indicators based on the UV luminosity or IR and far-IR
luminosities, and \SI{1}{Gyr} for indicators based on $D_{\rm n}(4000)$ index. Note that we only present the SFR estimates for galaxies that are classified as star-forming, defined as those with $\log \mathrm{SFR} > \log M_* \times 0.76 - 8.1$. This threshold is derived by shifting the best-fit star-forming main sequence from \citet{Renzini_2015} downward by approximately \SI{0.5}{dex}.
Here one can see that
the standard deviation is about \SI{0.3}{dex} for the SFR smoothed on these three timescales.
Notably, the resolution limit of TNG100 ($\sim 10^6 M_\odot$ for baryonic mass) suggests a minimally detectable SFR of approximately \SI{0.1}{\textit{M}_\odot.yr^{-1}} over a 10Myr time scale for particle-based methods.

Finally, Figures \ref{fig:tng_tng_sfh} and \ref{fig:tng_tng_tx} show the comparison of star formation
histories. The SFHs are parameterized with the method described in Appendix \ref{app:sfh_binning_test}. From the SFH residual shown in Figure~\ref{fig:tng_tng_sfh}, the SFHs of individual galaxies can be recovered within \SI{0.1}{dex} when the median is considered, and within \SIrange{0.3}{0.4}{dex} error when $1 \sigma$ distribution is considered for both low-mass and high-mass galaxies. Note that the residual distribution of high-mass galaxies (red) is slightly broader than that of low-mass galaxies, possibly due to the better coverage of SFH space for low-mass galaxies since they are more abundant in simulations.
Here, we also represent the SFH with the cosmic time
that a certain fraction ($50\%$, $70\%$, and $90\%$) of stars have formed in Figure~\ref{fig:tng_tng_tx}.
Again, it is evident that our method can recover the SFH for
individual galaxies quite well with scatter below \mbox{$\sim$}\SI{0.2}{dex}.

\section{The application to SDSS}%
\label{sec:applications_on_sdss}

Following successful validation on the IllustrisTNG simulation, we apply our method to the SDSS main galaxy sample to infer stellar mass, current SFR, and SFH from broadband photometry, utilizing the TNG100 galaxies as templates for mass-to-light ratio, SFH, and metal enrichment history. 

\subsection{Stellar mass and SFR}%
\label{sub:tng_sdss_msfr}

\begin{figure*}
    \centering
    \includegraphics[width=0.75\linewidth]{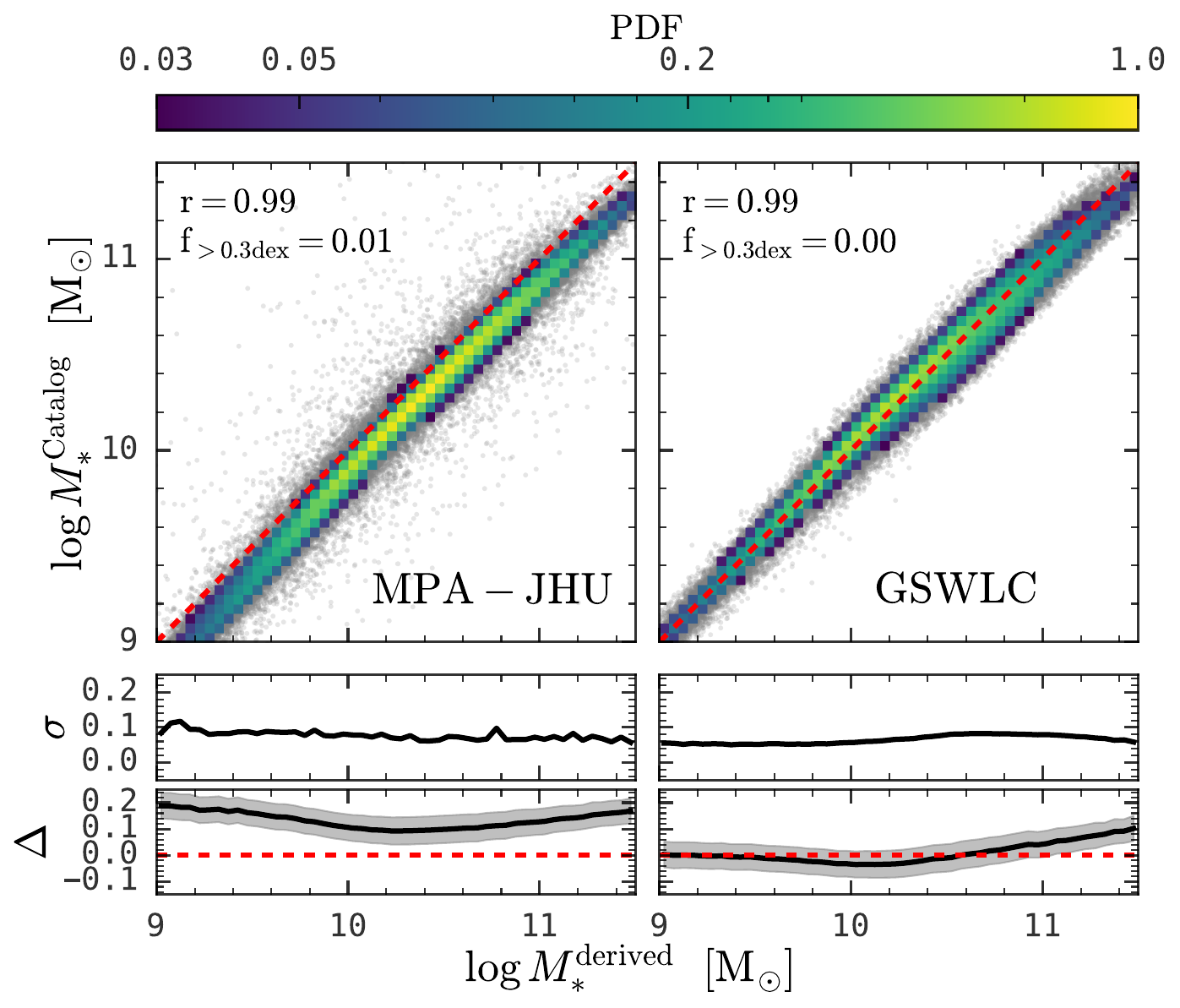}
    \caption{
        The top panels show the comparison of the stellar mass estimated by our
        method versus those in MPA-JHU
        \citep[left;][]{Kauffmann_2003} and GSWLC \citep[right;][]{Salim_2016}. Individual
        galaxies in SDSS are represented by gray dots, and the heat map illustrates the normalized probability distribution.
        The bottom panels show the standard deviations and mean differences as functions of estimated stellar mass. Our method aligns closely with GSWLC values and consistently shows an approximate \qty{0.13}{dex} systematic increase over MPA-JHU estimates.
    }%
    \label{fig:tng_sdss_mstar}
\end{figure*}

\begin{figure*}
    \centering
    \includegraphics[width=0.75\linewidth]{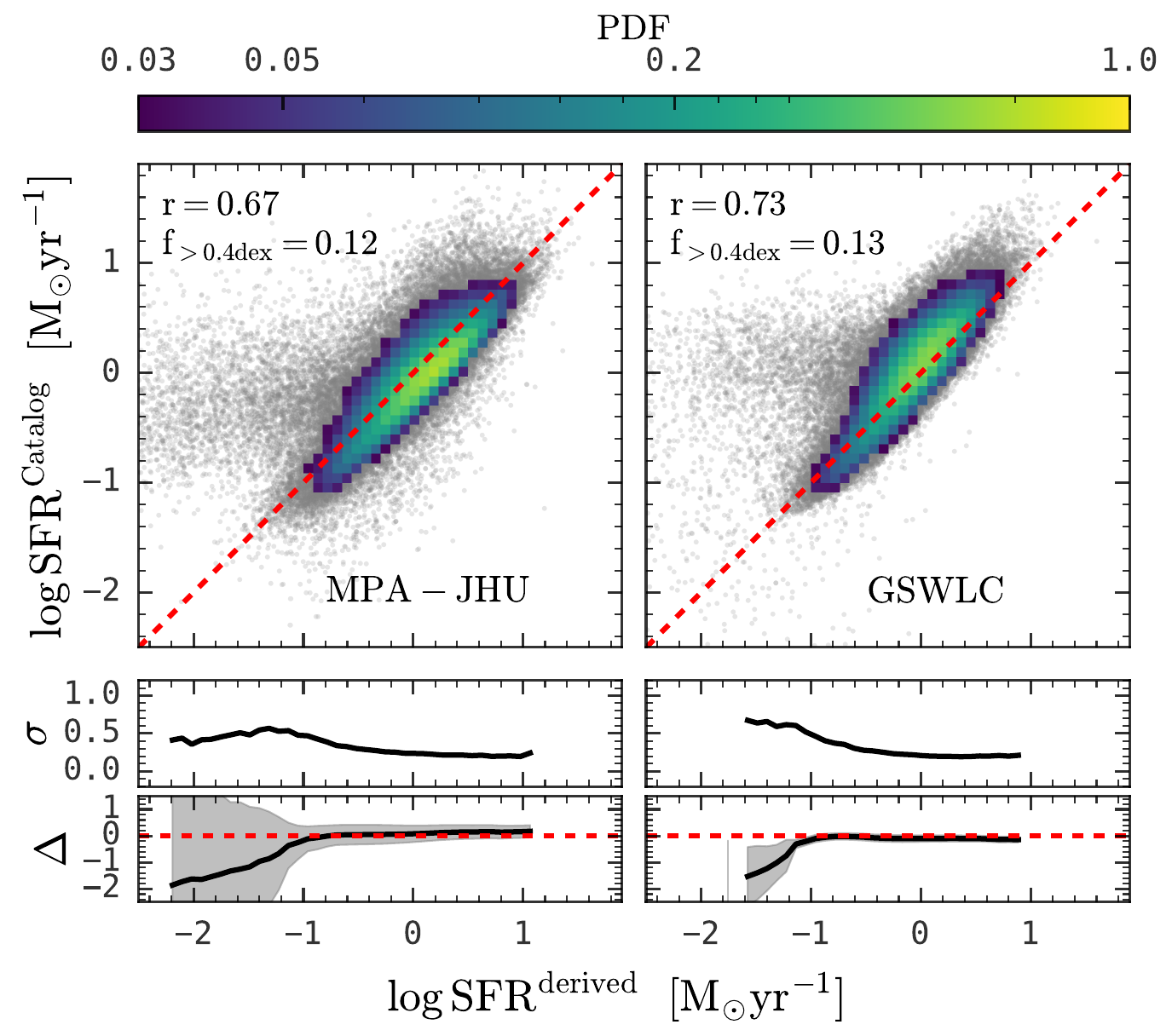}
    \caption{
        Similar to Figure~\ref{fig:tng_sdss_mstar}, focusing on the SFR for star-forming galaxies. Individual star-forming SDSS
        galaxies are represented by gray dots, and the heat map illustrates the normalized probability distribution. Our estimates align well with MPA-JHU and GSWLC values above $0.1  M_{\odot} \rm yr^{-1}$.
    }
    \label{fig:tng_sdss_sfr}
\end{figure*}

\begin{figure*}
    \centering
    \includegraphics[width=0.9\linewidth]{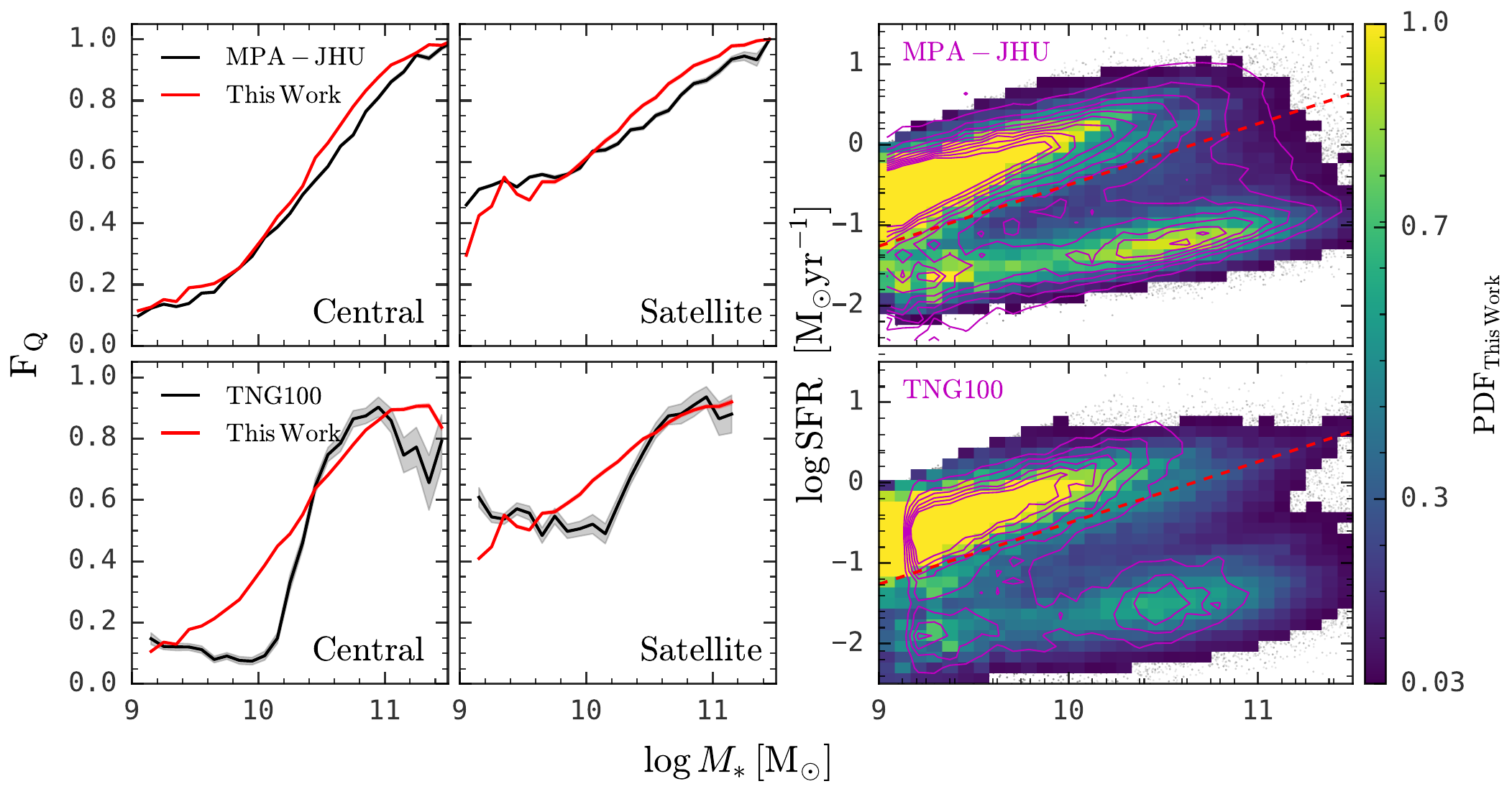}
    \caption{
        The left four panels show the quiescent galaxies fraction as a
        function of stellar mass for central (left) and satellite (right)
        galaxies. The red solid lines are our estimates smoothed over \qty{10}{Myr} (top) and \qty{100}{Myr} (bottom), in comparison with
        the results from the MPA-JHU catalog (top) and the TNG100
        simulation (bottom). The right two panels show the distribution of
        galaxies on the $\log M_* \rm - \log SFR$ plane. Individual
        galaxies are represented by gray dots, and the heat map illustrates the normalized probability distribution. The red
        dashed line ($\log \mathrm{SFR} = 0.76 \times \log M_* -8.1$) separates the
        star-forming and quiescent galaxies. The magenta solid contour lines
        are for our results with the SFR smoothed on 10 Myr
        (top) and 100 Myr (bottom), and the color maps are for the MPA-JHU
        result and the TNG100 simulation, respectively. Our method's results closely match the MPA-JHU observations in bimodality and show distinct deviations from TNG100 predictions, reinforcing the effectiveness of using TNG100 as a prior without replicating simulation biases.
}
    \label{fig:tng_sdss_msfr}
\end{figure*}

Figure~\ref{fig:tng_sdss_mstar} shows the stellar mass recovered with our method
in comparison with values in the MPA-JHU and GSWLC catalog. Note that the
MPA-JHU catalog derives stellar mass from the spectral indices for the central
part covered by the fiber and from the broadband photometry for the outer
part. Our result agrees with that of the MPA-JHU with the standard deviation
$\lesssim 0.1$ dex, except that our stellar mass is systematically larger than
theirs by about 0.13 dex. Note that we already corrected the difference caused
by the initial mass function and cosmology. Such a systematic difference is also reported by
\citet{Salim_2016}, where they attribute it to the different parametric
function forms for the SFH. From the right panels of
Figure~\ref{fig:tng_sdss_mstar}, one can see that our result agrees with the
GSWLC catalog \citep{Salim_2016} quite well with a standard deviation $\lesssim
0.1$ dex and negligible systematics, which suggests that the stellar mass in
MPA-JHU is indeed biased by their nonphysical priors for SFH.
Moreover, it is noteworthy that the GSWLC catalog derives the stellar mass from
the UV-to-IR photometry, while we only use five optical broad bands of
photometry.

Figure~\ref{fig:tng_sdss_sfr} shows the comparison of SFR for
star-forming galaxies between our
estimation and that of the MPA-JHU and GSWLC catalogs. For this figure, the star-forming galaxies are selected following the color-magnitude diagram criteria from \citet{Baldry_2004} since there is no real SFR in observation. We also test the results with different criteria, including the color--stellar mass diagram \citep{Cui_2024} and stellar mass--SFR diagram ($\log \mathrm{SFR} > \log M_* \times 0.76 - 0.81$, galaxies above the red dashed line in the upper right panel of Figure~\ref{fig:tng_sdss_msfr}). There are no noticeable changes among these different definitions. For passive galaxies, their SFRs in MPA-JHU and GSWLC have high uncertainties \citep{Salim_2016} and may even be upper limits, possibly due to the inaccurate photometry for GSWLC \citep{Li_2023} and uncertainty in the 
$D_{\rm n}4000$--sSFR relation \citep{Brinchmann_2004}. So, these passive galaxies are not considered for comparison.
Note that the star
formation rate in the MPA-JHU catalog is calibrated with the $\rm H\alpha$
luminosity, which traces the star formation activity in the last 10Myr, and
that of GSWLC is estimated by smoothing the inferred SFH on
100Myr. Thus, we choose to take the averaged SFR over 10Myr and
100Myr for a fair comparison with each catalog. Here one can see that, for
galaxies with SFR $>$ \SI{0.1}{\textit{M}_\odot.yr^{-1}}, our method produces consistent
results with MPA-JHU and GSWLC with the standard deviation about \SIrange{0.2}{0.5}{dex}
and negligible systematics. Finally, it should be emphasized that the GSWLC
catalog uses additional information from the UV and IR emissions, and MPA-JHU
uses the $\rm H\alpha$ emission and $D_{\rm n}(4000)$ spectral index, while
our method only uses five optical broad bands of photometry. Still, we yield
consistent results, indicating the superiority of our method.

We note that, for some galaxies, our model systematically provides lower SFR estimates than those from MPA-JHU and GSWLC. Although the true SFRs are unknown, these galaxies are all classified as star-forming; therefore, the measurements from MPA-JHU and GSWLC, which utilize $\rm H \alpha$ emission and UV plus IR photometry, respectively, should theoretically be more accurate. It is also interesting to note that this discrepancy is almost nonexistent in the self-consistency test (Figure~\ref{fig:tng_tng_sfr}), suggesting that the issue likely does not originate from the link between SEDs and SFH or our method itself. One possible explanation is the insufficient sampling of SFH space in the simulation. However, this problem does not significantly impact the results statistically, as the catastrophic failure rate is low (approximately 20\%).

Before we step into the comparison of quiescent fraction and SFR bimodality,
several technical considerations should be mentioned. Firstly, we choose to
compare with the MPA-JHU catalog rather than the GSWLC catalog, since the
latter suffers from incompleteness due to the requirement of cross-matching
with the UV and IR photometry catalog. Secondly, we note that TNG100 produces
many quiescent galaxies with SFR equal to zero, which is not suitable for
presentation, so we assign some small SFR to these galaxies for the sake of
fair comparison with observational results. It is noteworthy that some work argues that the distribution of galaxies on the $\log M_* \rm  - \log SFR$ plane is unimodal: a star-forming peak and an extended tail toward low SFR \citep[see][]{Renzini_2015, Feldmann_2017, Eales_2017}. Thirdly, the calculation of SFR from the amount of stellar particles formed in
the last period suffers from the shot noise. We note that the baryon
cell weighs about $10^6 M_\odot$ in TNG100, so the minimal SFR above zero in
the last 10Myr will be \SI{0.1}{\textit{M}_\odot.yr^{-1}}, which corresponds to a specific
SFR of $10^{-10}\,\si{yr^{-1}}$
for a $10^9 M_\odot$  galaxy and, hence, it will
affect the calculation of the quiescent fraction. Therefore, we choose to smooth
the SFR over 100Myr to minimize this effect when we compare our
results with TNG100.

Figure~\ref{fig:tng_sdss_msfr} shows the quiescent fraction for central and
satellite galaxies, respectively, and the apparent bimodal distribution of galaxies on
the $\log M_* \rm  - \log SFR$ plane. Here, we employ the group catalog in \citet{Yang_2007}
to classify central and satellite galaxies. Firstly, the quiescent fraction for
central galaxies monotonically increases with stellar mass from about 10\% at
$10^9 M_\odot$ to almost 100\% at $10^{11.5} M_\odot$, which is a manifestation of mass quenching. As reported in \citet{Wang_2023}, a large portion of low-
mass quiescent central galaxies are backsplash galaxies, which were previously
satellite galaxies of other massive halos and got ejected out of their host
halos. Secondly, the quiescent fraction of the central galaxies in TNG100
abruptly increases around $10^{10.2} M_\odot$, which is absent in observational
data. The sharp transition may result from the problematic active galactic nucleus (AGN) feedback interpretation in TNG \citep[see][for more information]{Terrazas_2020}.
In addition, the result obtained from our method is consistent with
MPA-JHU and previous results \citep{Baldry_2006, Peng_2010, Peng_2012, Wetzel_2013, Trussler_2020, Gallazzi_2021} while significantly differing from the TNG100 simulation, thus confirming the robustness of our approach in applying physical priors without duplicating simulation results. Thirdly, satellite galaxies are more quenched than
central galaxies at a given stellar mass, since they are additionally subject to
environmental effects. Finally, we found three components in the distribution
of galaxies on the $\log M_* \rm -\log SFR$ plane: one star-forming main sequence, one
massive quiescent component contributed by mass quenching, and one low-mass
quiescent component from environmental quenching
\citep[see][]{Renzini_2015}. One notable thing is that we
can accurately reproduce the star-forming main sequence exhibited in the
MPA-JHU catalog, which is derived from $\rm H\alpha$ luminosity, while our
results are only based on five broad bands of photometry.

\subsection{Cosmic SFR density and stellar mass density}%
\label{sub:estimating_star_formation_histories}

\begin{figure*}
    \centering
    \includegraphics[width=0.9\linewidth]{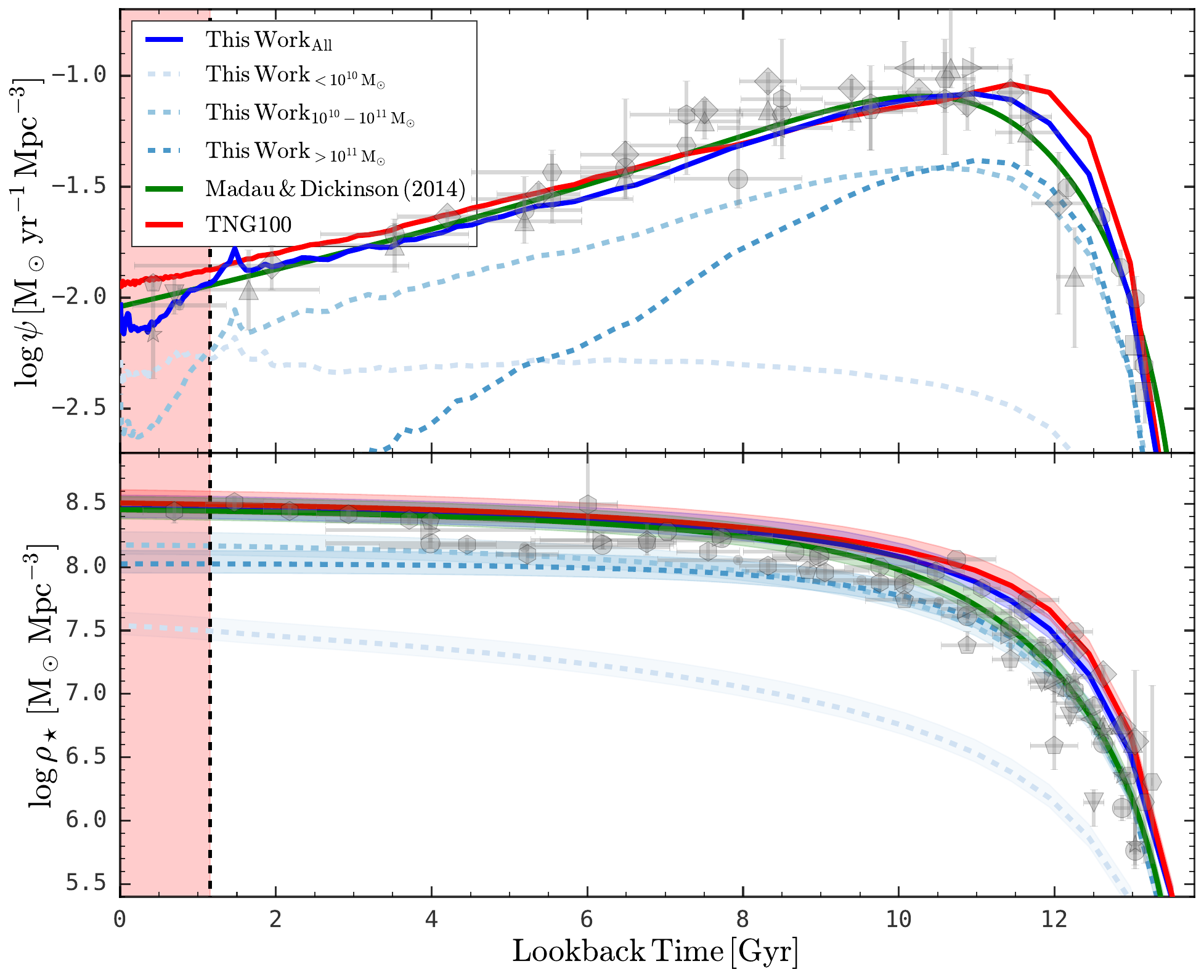}
    \caption{
        The upper panel depicts the comparison of the cosmic SFR density, defined
        as the comoving SFR density as a function of look-back time. 
        The gray symbols are the observational results from direct SFR density estimates at different redshifts \citep{Sanders_2003, Takeuchi_2003, Wyder_2005, Schiminovich_2005, Dahlen_2007, Reddy_2009,
        Robotham_2011, Magnelli_2011, Cucciati_2012, Bouwens_2012a, Bouwens_2012b, Schenker_2013,  Magnelli_2013, Gruppioni_2013, Madau_2014}, and the fitting line of these
        observational results is shown as the green solid line
        \citep{Madau_2014}. The result of the TNG100 simulation is shown in
        the red solid line and the result obtained with our method is
        shown in the blue solid line. The three blue dashed lines are the
        cosmic SFR density that end up in descendant galaxies
        with different stellar masses at $z\sim 0$. 
        The lower panel shows the comparison of the cosmic stellar mass density, defined
        as the comoving stellar mass density as a function of look-back time.
        Solid lines are integrated from cosmic SFR density with a return factor of 0.41 (based on the Chabrier initial mass function). The error bands are based on return factors of 0.25 and 0.5. The gray symbols are the observational results integrated from stellar mass functions between $10^8 M_{\odot}$ and $10^{13} M_{\odot}$ \citep{Caputi_2011, Gonzalez_2011, Santini_2012, Ilbert_2013, Muzzin_2013, Duncan_2014,  Grazian_2015, Caputi_2015, Wright_2018, Kikuchihara_2020, McLeod_2021, Thorne_2021, Weaver_2023}.
        Our method effectively captures both the cosmic SFR density and cosmic stellar mass density, aligning closely with direct measurements and revealing detailed insights into galaxy evolution. 
    }
    \label{fig:tng_sdss_sfrd}
\end{figure*}

As demonstrated in \S\,\ref{sec:method} and tested in
\S\,\ref{sec:testing_on_tng100}, our method is able to recover the star
formation history of each galaxy from their broadband photometry.
By doing so for all galaxies in the SDSS main galaxy sample, the
cosmic SFR density, i.e., the comoving SFR density as
a function of redshift, can be recovered by aggregating individual SFHs of local
galaxies and dividing it by the comoving volume occupied by these galaxies. This is known as the fossil record method and has been applied in many works \citep{Heavens_2004, Gallazzi_2008, Madau_2014, Carnall_2019a, Leja_2019a, Leja_2019b, Sanchez_2019}, which should have similar results to direct observation \citep{Lilly_1996, Madau_1996a} in principle.
It is noteworthy that, since SDSS is a flux-limited sample, the comoving volume
occupied by observed galaxies is dependent on the intrinsic luminosity, which
is commonly refer to as $V_{\rm max}$ \citep{Schmidt_1968}.

The top panel of Figure~\ref{fig:tng_sdss_sfrd} shows the cosmic SFR density
reconstructed from the individual 100 bins of SFH of local galaxies in the
blue solid line. For comparison, we also present the result from TNG100 in the red
solid line, and the direct observational measurement results in gray symbols with
the fitting function in the green solid line \citep{Sanders_2003, Takeuchi_2003, Wyder_2005, Schiminovich_2005, Dahlen_2007, Reddy_2009,
Robotham_2011, Magnelli_2011, Cucciati_2012, Bouwens_2012a, Bouwens_2012b, Schenker_2013,  Magnelli_2013, Gruppioni_2013, Madau_2014}. Here, one can see that TNG100
slightly over-predicts the peak position by $\approx \!\SI{1}{Gyr}$
, while our
method, although using TNG100 as a prior, is a better match to the observational result from direct measurements. It is noteworthy that our method outperforms several previous parametric and nonparametric SFH
methods \citep[see][]{Carnall_2019a, Leja_2019b}, which confirms the superiority
of our method in delivering physical parameters from the multiband
photometry.

The three dashed lines in the top panel of Figure~\ref{fig:tng_sdss_sfrd} show the cosmic star
formation histories contributed by $z \sim 0$ descendant galaxies with different stellar
masses. For massive galaxies, although their assembly is relatively recent, most
of their stars were already formed more than \SI{10}{Gyr} ago, which is known as the
archeology downsizing effect \citep{Thomas_2005, Nelan_2005, Heavens_2004, Jimenez_2005, Neistein_2006, Renzini_2006}. Meanwhile, stars formed in the recent several gigayears
end up in low-mass galaxies, which is expected since massive galaxies are
quenched and stop producing new stars (see Figure~\ref{fig:tng_sdss_msfr}).

By integrating the cosmic SFR density and assuming a return factor of 0.41 from the Chabrier initial mass function, we also derive the cosmic stellar mass density in the bottom panel of Figure~\ref{fig:tng_sdss_sfrd}. 
The results from direct SFR density estimates at different
redshifts \citep{Madau_2014} and TNG100 are also shown in green and red for comparison.
Gray symbols are cosmic stellar mass density integrated from stellar mass function in the literature \citep{Caputi_2011, Gonzalez_2011, Santini_2012, Ilbert_2013, Muzzin_2013, Duncan_2014,  Grazian_2015, Caputi_2015, Wright_2018, Kikuchihara_2020, McLeod_2021, Thorne_2021, Weaver_2023}. Our results agree with both \citep{Madau_2014} and stellar mass function results well, while TNG100 slightly exceeds most measurements at about \SI{10}{Gyr} ago. When categorizing galaxies into descendant stellar mass bins, we again see the archeology downsizing effect: massive galaxies form their mass early and gradually cease star formation. By contrast, low-mass galaxies keep on creating new stars.

\subsection{Reconstructing spectral indices}
\label{subsec:spec_idx}
\begin{figure*}
    \centering
    \includegraphics[width=0.7\linewidth]{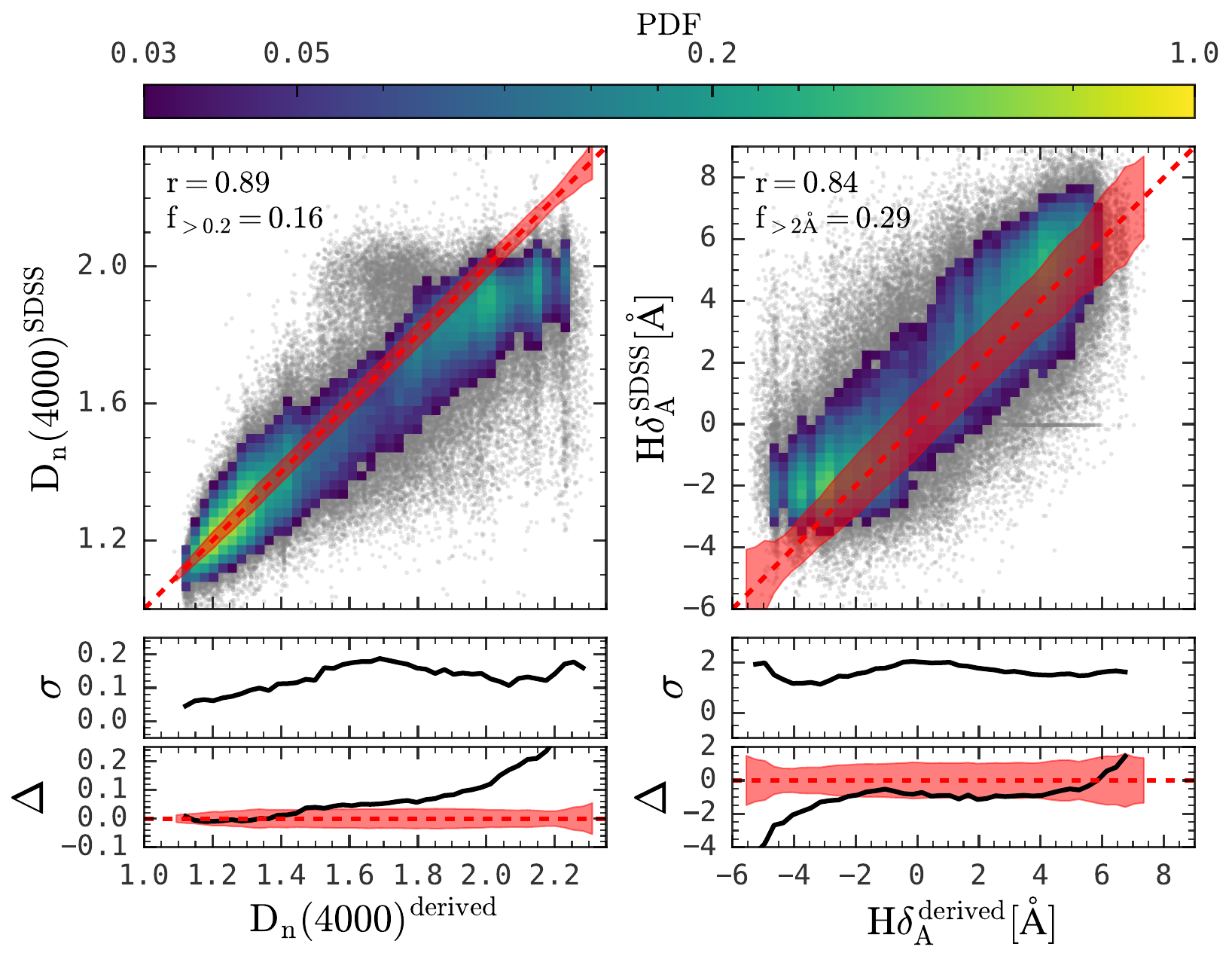}
    \caption{
        The top panels show the comparison between measured and reconstructed spectral indices, $D_{\rm n}(4000)$ (left) and $\rm H\delta_A$ (right) for the SDSS main galaxy sample. Individual
        galaxies are represented by gray dots, and the heat map illustrates the normalized probability distribution. The bottom panels highlight the standard deviation and systematic bias relative to directly measured values.
        The red-shaded regions show the median observational errors.
        This figure shows that our
        reconstructed spectral indices resemble the directly measured values
        with strong correlations (Spearman's coefficient $\approx 0.85$)
    }%
    \label{fig:tng_sdss_idx}
\end{figure*}

Previous studies found that the spectral indices of observed galaxies contain
abundant information about the physical properties of galaxies, and these
spectral indices are used to derive physical properties, such as stellar mass,
SFH, and stellar metallicity \citep{Faber_1972, Worthey_1997, Kauffmann_2003, Gallazzi_2005}, and
are incorporated into the full-spectrum fitting procedure to deliver more
accurate property estimations. Thus, given the success of our method based on
broadband photometry, it is pertinent to investigate the extent to which the spectral indices of observed galaxies can be reconstructed. Here, we focus on $D_{\rm n}(4000)$ and $\rm
H\delta_A$. Here, $D_{\rm n}(4000)$ quantifies the discontinuity around 4000\AA, which comes from the absorption of ionized metals, and it is defined as
the ratio of the average flux density in the bands 3850--3950 and \SIrange{4000}{4100}{\angstrom}
\citep[see][]{Balogh_1999, Kauffmann_2003}. $D_{\rm n}(4000)$ is profoundly
dependent on the stellar age and stellar metallicity: older and more metal-rich
galaxies exhibit higher $D_{\rm n}(4000)$. Particularly, $D_{\rm n}(4000)$ is sensitive
to the median stellar age. $\rm H\delta_A$
quantifies the $\rm H\delta$ absorption lines \citep{Worthey_1997}, which are mainly contributed by
late-B and early-F stars, and therefore $\rm H\delta_A$ is a sensitive probe
of star formation activities that occurred \SIrange{0.1}{1}{Gyr} ago \citep{Kauffmann_2004}. Notice that the $D_{\rm n}(4000)$ and $\rm H\delta_A$ distributions of TNG100 galaxies are similar to observations \citep{Wu_2018, Wu_2021}, indicating that SFHs in TNG produce galaxy properties consistent with the observations. It is noteworthy that both
spectral indices are not significantly affected by the presence of dust absorption \citep[][also see the test in Appendix~\ref{app:spec_idx_test}]{Kauffmann_2003, MacArthur_2005}, allowing for the safe exclusion of this effect in the forward synthesis of these two spectral indices.

Here is the procedure to reconstruct the spectral indices ($D_{\rm n}(4000)$ and
$\rm H\delta_A$) from the broadband photometry of observed SDSS galaxies.
Firstly, we derive the SFH and the metal enrichment history of
each observed galaxy from its broadband photometry using our method (see
\S\,\ref{sec:method}) with TNG100 as a physical prior. Secondly, we synthesize
the full spectrum of each galaxy from its SFH and
metal enrichment history using the \textsc{fsps} package adopting the MILES
stellar library, Padova isochrones, and Chabrier initial mass function. Finally, we measure the
spectral indices following the procedure done for the observed spectrum.

Figure~\ref{fig:tng_sdss_idx} shows the comparison between the reconstructed and
the observed spectral indices. It should be noted that we exclude galaxies with
unreliable spectral index measurements, which includes galaxies whose
measurement error in $D_{\rm n}(4000)$ is above 0.07 or whose measurement error in
$\rm H\delta_A$ is above 2.5\AA. This excludes 5\% and 4\% of SDSS
galaxies, respectively. Here, one can see that Spearman's correlation
coefficients for both indices are around 0.85, which means that about 85\% of
the variation for these two spectral indices can be captured by the optical
broadband photometry. Besides, the reconstructed spectral indices closely
follow the observed values on the one-to-one line except for galaxies with high
$D_{\rm n}(4000)$ and negative $\rm H\delta_A$, which are quenched galaxies and
dominated by old stellar populations. We have inspected the result for
star-forming galaxies (not shown here) and found that the systematic deviation
becomes negligible.

Many previous studies have conducted similar tests on different SED-fitting procedures. For instance, \citet{Chen_2012} developed a method based on principal component analysis (PCA) and applied it to BOSS spectra. They were able to retrieve the $D_{\rm n}(4000)$ and $\rm H\delta_A$ indices for high-redshift galaxies ($0.4 < z < 0.7$) with a precision comparable to our model, benefiting from the rich information available in the full spectra. In a related study, \citet{Nersesian_2024} utilized the Prospector procedure on galaxies around $z \sim 1$ with the COSMOS2020 photometry. Their results, which involved comparisons with direct observations from the Large Early Galaxy Astrophysics Census (LEGA-C) survey \citep{vanderWel_2016, vanderWel_2021}, demonstrated that the recovered $\rm H\delta_A$ indices generally aligned with actual measurements, showing no significant systematic offsets. These findings are consistent with our results.

The success in reconstructing spectral indices has three
implications. Firstly, our method can deliver physical SFHs
and metal enrichment histories, so that the synthesized spectral indices resemble
the results obtained from direct observation, even though we only use
broadband photometry. Secondly, much information provided by these two
spectral indices is contained in broadband photometry. Thirdly, the
remaining $\approx \! \qty{15}{\percent}$ variation contains additional information about the
underlying interesting physical properties \citep{Nersesian_2024}, whose extraction requires an extension of
the current method to the full-spectrum situation. This could be explored in future work.

\section{Summary}
\label{sec:conslusions}

The SEDs of galaxies contain rich information about their evolutionary histories and have been routinely used to derive fundamental galaxy properties such as stellar mass, SFR, stellar age, and star formation history (SFH) through SED-fitting techniques. However, SED fitting, especially when limited to broadband photometry, may not be sufficient to tightly constrain the SFH and metal enrichment history. Consequently, conventional methods, both parametric and nonparametric, often fail to accurately recover the observed cosmic SFR density, due to strong degeneracy, and oversimplified and nonphysical assumptions about SFHs that are unable to account for complex SFH variations across different galaxy populations.

To address this issue, this study introduces a novel method for estimating the physical properties of galaxies from their observed broadband SEDs, utilizing hydrodynamical galaxy formation simulations as priors. We validated this method on simulated galaxies and subsequently applied it to observations. Our main findings are summarized as follows:

\begin{enumerate}

        \item Testing on the TNG100 SDSS-mock with five-band photometry ($u, g, r, i, z$) including dust effects, our method accurately recovers the stellar mass, current SFR, and typical galaxy formation times from optical broadband photometry. The standard deviations are around 0.05, 0.3, and \qty{0.2}{dex}, respectively, and the systematic bias is negligible (see
        Figure~\ref{fig:tng_tng_mstar}, \ref{fig:tng_tng_sfr}, and
        \ref{fig:tng_tng_tx}).

        \item When applied to observed SDSS galaxies using optical photometry ($u, g, r, i, z$) with spectroscopic redshift, our method consistently estimates stellar mass and SFRs that align with those in the MPA-JHU and GSWLC catalogs, showing a standard deviation of \mbox{$\sim$}\SI{0.1}{dex} for stellar mass and \mbox{$\sim$}\SI{0.4}{dex} for SFR (see Figure~\ref{fig:tng_sdss_mstar} and \ref{fig:tng_sdss_sfr}).

    \item Using the derived stellar mass and SFR from our method on a mass-completed sample in SDSS, the bimodal distribution of galaxies on the $\log M_* \rm -\log SFR$ plane (i.e., main sequence and passive sequence) and the quenched fraction for centrals and satellites all closely match those obtained from MPA-JHU results. \textit{Interestingly, despite employing SFHs from TNG100 as priors, the statistical properties of derived galaxy properties (such as the main sequence, quenched fraction, and environmental effects) are more aligned with MPA-JHU than TNG itself} (see Figure~\ref{fig:tng_sdss_msfr}). For instance, the sharp quenching threshold around $\log M_* \sim 10.5$ in TNG100 (as shown in Figure~\ref{fig:tng_sdss_msfr}) is due to its implemented AGN feedback model. The quenched fractions derived from our method evidently have not been influenced by these specific quenching mechanisms in TNG.

    \item Using the SFHs of individual SDSS galaxies derived from our method, we estimate the cosmic SFR density, and also the cosmic stellar mass density by integrating these SFHs. Both results show remarkable consistency with direct observational measurements up to $z \sim 6$. This is the first time that SFHs derived from SEDs accurately match direct observational measurements. We also find that the stellar populations of massive galaxies are already largely formed about \qty{10}{Gyr} ago, while recent star formation predominantly occurs in low-mass galaxies (see Figure~\ref{fig:tng_sdss_sfrd}), consistent with the downsizing trend established in the literature.

    \item From the estimated SFHs and metal enrichment histories of
        individual galaxies in SDSS, we synthesize two spectral indices, $\rm
        D_n(4000)$ and $\rm H\delta_A$.
        Our synthesized spectral indices closely match direct observational results. This validation supports the reliability of our derived star formation and metal enrichment histories, despite solely relying on broadband photometry (see Figure~\ref{fig:tng_sdss_idx}).

\end{enumerate}

It is important to note that our method does not depend on simulations being considered ``the truth'' (e.g. TNG100). Unlike conventional parametric methods, simulations offer a large set of physically realistic SFH templates. As a result, \textit{our approach effectively excludes many physically unrealistic SFHs that are often included in conventional SED fitting methods, which might otherwise yield better fits due to strong degeneracies within the SED.} This exclusion of physically unrealistic templates represents a significant advantage of our methodology.

Meanwhile, we must also emphasize that \textit{galaxies in the real universe may or may not follow every SFH provided by the simulations. And the relative fraction of certain galaxy populations that evolve according to a specific SFH may differ significantly from those predicted by the simulations.} In fact, when applying our method to SDSS galaxies, as illustrated in Figure~\ref{fig:tng_sdss_msfr}, we successfully recovered the observed quench fraction and the detailed distribution of galaxies on the $\log M_* \rm  - \log SFR$ plane, whereas TNG100 produced noticeably different results. The derived star formation rate density and stellar mass density from our method also show a better match with direct observations compared to TNG100. Again, we emphasize that it is not necessary for the TNG simulations to be an accurate representation of reality.
They just offer an extremely broad set of plausible SFHs, much wider than any prior parametric or nonparametric
approach.

Our method not only provides robust and accurate estimates of stellar population evolution, particularly galaxy SFHs, using only broadband photometry but also demonstrates high efficiency. Our tests show that it can process the star formation and metal enrichment histories for one million galaxies in less than 10 minutes using a single Intel-i5 core. These features make it an exceptionally promising tool for ongoing facilities like Euclid \citep{Laureijs_2011} and upcoming ones such as the Chinese Space Station Telescope \citep{Zhan_2011} and the
Large Synoptic Survey Telescope \citep{Ivezic_2019}, which will deliver vast quantities of high-quality multiband photometric data. Moreover, its outstanding computational efficiency allows for the application of pixel-to-pixel SED fittings to derive the resolved galaxy properties. In the future, we will use this method also for galaxy samples at high redshifts, using TNG templates at the corresponding cosmic times.

\section*{Acknowledgements}
We gratefully acknowledge Dr. John R. Weaver for providing the data from \citet{Weaver_2023} and offering helpful explanations. Y.P. and Z.G. acknowledge the support from the National Key R\&D Program of China (2022YFF0503401),  the National Science Foundation of China (NSFC) grant Nos. 12125301, 12192220, 12192222, and the science research grants from the China Manned Space Project with No. CMS-CSST-2021-A07. Y.P. acknowledges support from the New Cornerstone Science Foundation through the XPLORER PRIZE. L.C.H. acknowledges the National Science Foundation of China (11991052, 12233001), the National Key R\&D Program of China (2022YFF0503401), and the China Manned Space Project (CMS-CSST-2021-A04, CMS-CSST-2021-A06). J.D. acknowledges the support from the National Science Foundation of China (NSFC) grant No. 12303010. Q.G. is supported by the National Natural Science Foundation of China (Nos. 12192222, 12192220 and 12121003)

This work is extensively supported by the High-performance Computing Platform of Peking University, China.
The authors acknowledge the Tsinghua Astrophysics High-Performance
Computing platform at Tsinghua University for providing computational and data
storage resources that have contributed to the research results reported within
this paper.

This research made use of NASA’s Astrophysics Data System for bibliographic information.
The computation in this work is supported by the HPC toolkit \textsc{HIPP} \citep{Chen_2023} \footnote{\url{https://github.com/ChenYangyao/hipp}}. 
\software{\textsc{ipython} \citep{Perez_2007},
\textsc{numpy} \citep{Harris_2020},
\textsc{pandas} \citep{ThepandasdevelopmentTeam_2024},
\textsc{scipy} \citep{Virtanen_2020},
\textsc{astropy} \citep{AstropyCollaboration_2022},
\textsc{h5py} \citep{Collette_2023},
\textsc{matplotlib} \citep{Hunter_2007},
\textsc{numba} \citep{Lam_2015},
\textsc{seaborn} \citep{Waskom_2021},
\textsc{extinction} \citep{Barbary_2016},
\textsc{tqdm} \citep{daCosta-Luis_2024},}

\section*{Data availability}

The derived properties in this paper are publicly available on Zenodo under an open-source 
Creative Commons Attribution license at 
\dataset[doi:10.5281/zenodo.14209295]{https://doi.org/10.5281/zenodo.14209295}. and are described in Table~\ref{tab:cat}. The detailed star formation history and halo properties, including halo mass and halo assembly history, will be fully publicly
available in our following data release paper, together
with various other derived properties of the halos and
galaxies.

\begin{deluxetable*}{cc}
\tablecaption{Catalog Columns\label{tab:cat}}
\tablehead{
\colhead{Column name} & \colhead{Description} 
}
\startdata
\texttt{plate} & SDSS spectroscopic plate number \\
\texttt{fiberid}  & SDSS spectroscopic fiber number\\
\texttt{mjd} & SDSS spectroscopic observation date\\
\texttt{ra} & SDSS R.A. in J2000\\
\texttt{dec} & SDSS decl. in J2000\\
\texttt{z} & SDSS spectroscopic redshift\\
\texttt{log\_mstar} & Stellar Mass [$M_{\odot}$]\\
\texttt{log\_mstar\_err} & Error of the stellar mass [$M_{\odot}$]\\
\texttt{sfr\_10myr/100myr/1gyr} & Star formation rate averaged over the recent \qty{10}{Myr}/\qty{100}{Myr}/\qty{1}{Gyr}
[$M_{\odot} \rm yr^{-1}$] \\
\texttt{sfr\_10myr/100myr/1gyr\_err} & Error of star formation rate averaged over the recent \qty{10}{Myr}/\qty{100}{Myr}/\qty{1}{Gyr} timescale [$M_{\odot} \rm yr^{-1}$] \\
\texttt{t50/70/90} & Cosmic time that \qty{50}{\percent}/\qty{70}{\percent}/\qty{90}{\percent} of the mass have formed [Gyr] \\
\texttt{t50/70/90\_err} & Error of cosmic time that \qty{50}{\percent}/\qty{70}{\percent}/\qty{90}{\percent} of the mass have formed [Gyr] \\
\texttt{weighted\_dist} & Weighted distance $\mathcal{D}_i$ mentioned in \S\,\ref{sec:method}, galaxy with $\mathcal{D}_i \geq 0.19$ should not be used in principle\\
\enddata
\end{deluxetable*}

\bibliography{sfh_tng}{}

\begin{thebibliography}{}
\expandafter\ifx\csname natexlab\endcsname\relax\def\natexlab#1{#1}\fi
\providecommand{\url}[1]{\href{#1}{#1}}
\providecommand{\dodoi}[1]{doi:~\href{http://doi.org/#1}{\nolinkurl{#1}}}
\providecommand{\doeprint}[1]{\href{http://ascl.net/#1}{\nolinkurl{http://ascl.net/#1}}}
\providecommand{\doarXiv}[1]{\href{https://arxiv.org/abs/#1}{\nolinkurl{https://arxiv.org/abs/#1}}}

\bibitem[{{Abazajian} {et~al.}(2009){Abazajian}, {Adelman-McCarthy}, {Ag{\"u}eros}, {Allam}, {Allende Prieto}, {An}, {Anderson}, {Anderson}, {Annis}, {Bahcall}, \& et~al.}]{Abazajian_2009}
{Abazajian}, K.~N., {Adelman-McCarthy}, J.~K., {Ag{\"u}eros}, M.~A., {et~al.} 2009, \apjs, 182, 543, \dodoi{10.1088/0067-0049/182/2/543}

\bibitem[{{Abramson} {et~al.}(2016){Abramson}, {Gladders}, {Dressler}, {Oemler}, {Poggianti}, \& {Vulcani}}]{Abramson_2016}
{Abramson}, L.~E., {Gladders}, M.~D., {Dressler}, A., {et~al.} 2016, \apj, 832, 7, \dodoi{10.3847/0004-637X/832/1/7}

\bibitem[{{Astropy Collaboration} {et~al.}(2022){Astropy Collaboration}, {Price-Whelan}, {Lim}, {Earl}, {Starkman}, {Bradley}, {Shupe}, {Patil}, {Corrales}, {Brasseur}, {N{\"o}the}, {Donath}, {Tollerud}, {Morris}, {Ginsburg}, {Vaher}, {Weaver}, {Tocknell}, {Jamieson}, {van Kerkwijk}, {Robitaille}, {Merry}, {Bachetti}, {G{\"u}nther}, {Aldcroft}, {Alvarado-Montes}, {Archibald}, {B{\'o}di}, {Bapat}, {Barentsen}, {Baz{\'a}n}, {Biswas}, {Boquien}, {Burke}, {Cara}, {Cara}, {Conroy}, {Conseil}, {Craig}, {Cross}, {Cruz}, {D'Eugenio}, {Dencheva}, {Devillepoix}, {Dietrich}, {Eigenbrot}, {Erben}, {Ferreira}, {Foreman-Mackey}, {Fox}, {Freij}, {Garg}, {Geda}, {Glattly}, {Gondhalekar}, {Gordon}, {Grant}, {Greenfield}, {Groener}, {Guest}, {Gurovich}, {Handberg}, {Hart}, {Hatfield-Dodds}, {Homeier}, {Hosseinzadeh}, {Jenness}, {Jones}, {Joseph}, {Kalmbach}, {Karamehmetoglu}, {Ka{\l}uszy{\'n}ski}, {Kelley}, {Kern}, {Kerzendorf}, {Koch}, {Kulumani}, {Lee}, {Ly}, {Ma}, {MacBride}, {Maljaars}, {Muna}, {Murphy}, {Norman},
  {O'Steen}, {Oman}, {Pacifici}, {Pascual}, {Pascual-Granado}, {Patil}, {Perren}, {Pickering}, {Rastogi}, {Roulston}, {Ryan}, {Rykoff}, {Sabater}, {Sakurikar}, {Salgado}, {Sanghi}, {Saunders}, {Savchenko}, {Schwardt}, {Seifert-Eckert}, {Shih}, {Jain}, {Shukla}, {Sick}, {Simpson}, {Singanamalla}, {Singer}, {Singhal}, {Sinha}, {Sip{\H{o}}cz}, {Spitler}, {Stansby}, {Streicher}, {{\v{S}}umak}, {Swinbank}, {Taranu}, {Tewary}, {Tremblay}, {de Val-Borro}, {Van Kooten}, {Vasovi{\'c}}, {Verma}, {de Miranda Cardoso}, {Williams}, {Wilson}, {Winkel}, {Wood-Vasey}, {Xue}, {Yoachim}, {Zhang}, {Zonca}, \& {Astropy Project Contributors}}]{AstropyCollaboration_2022}
{Astropy Collaboration}, {Price-Whelan}, A.~M., {Lim}, P.~L., {et~al.} 2022, \apj, 935, 167, \dodoi{10.3847/1538-4357/ac7c74}

\bibitem[{{Baldry} {et~al.}(2006){Baldry}, {Balogh}, {Bower}, {Glazebrook}, {Nichol}, {Bamford}, \& {Budavari}}]{Baldry_2006}
{Baldry}, I.~K., {Balogh}, M.~L., {Bower}, R.~G., {et~al.} 2006, \mnras, 373, 469, \dodoi{10.1111/j.1365-2966.2006.11081.x}

\bibitem[{{Baldry} {et~al.}(2004){Baldry}, {Glazebrook}, {Brinkmann}, {Ivezi{\'c}}, {Lupton}, {Nichol}, \& {Szalay}}]{Baldry_2004}
{Baldry}, I.~K., {Glazebrook}, K., {Brinkmann}, J., {et~al.} 2004, \apj, 600, 681, \dodoi{10.1086/380092}

\bibitem[{{Balogh} {et~al.}(1999){Balogh}, {Morris}, {Yee}, {Carlberg}, \& {Ellingson}}]{Balogh_1999}
{Balogh}, M.~L., {Morris}, S.~L., {Yee}, H.~K.~C., {Carlberg}, R.~G., \& {Ellingson}, E. 1999, \apj, 527, 54, \dodoi{10.1086/308056}

\bibitem[{{Barbary}(2016)}]{Barbary_2016}
{Barbary}, K. 2016, {extinction v0.3.0},  Zenodo, \dodoi{10.5281/zenodo.804967}

\bibitem[{{Behroozi} {et~al.}(2019){Behroozi}, {Wechsler}, {Hearin}, \& {Conroy}}]{Behroozi_2019}
{Behroozi}, P., {Wechsler}, R.~H., {Hearin}, A.~P., \& {Conroy}, C. 2019, \mnras, 488, 3143, \dodoi{10.1093/mnras/stz1182}

\bibitem[{{Blanton} \& {Roweis}(2007)}]{Blanton_2007}
{Blanton}, M.~R., \& {Roweis}, S. 2007, \aj, 133, 734, \dodoi{10.1086/510127}

\bibitem[{{Blanton} {et~al.}(2005){Blanton}, {Schlegel}, {Strauss}, {Brinkmann}, {Finkbeiner}, {Fukugita}, {Gunn}, {Hogg}, {Ivezi{\'c}}, {Knapp}, {Lupton}, {Munn}, {Schneider}, {Tegmark}, \& {Zehavi}}]{Blanton_2005}
{Blanton}, M.~R., {Schlegel}, D.~J., {Strauss}, M.~A., {et~al.} 2005, \aj, 129, 2562, \dodoi{10.1086/429803}

\bibitem[{{Boquien} {et~al.}(2019){Boquien}, {Burgarella}, {Roehlly}, {Buat}, {Ciesla}, {Corre}, {Inoue}, \& {Salas}}]{Boquien_2019}
{Boquien}, M., {Burgarella}, D., {Roehlly}, Y., {et~al.} 2019, \aap, 622, A103, \dodoi{10.1051/0004-6361/201834156}

\bibitem[{{Bouch{\'e}} {et~al.}(2010){Bouch{\'e}}, {Dekel}, {Genzel}, {Genel}, {Cresci}, {F{\"o}rster Schreiber}, {Shapiro}, {Davies}, \& {Tacconi}}]{Bouche_2010}
{Bouch{\'e}}, N., {Dekel}, A., {Genzel}, R., {et~al.} 2010, \apj, 718, 1001, \dodoi{10.1088/0004-637X/718/2/1001}

\bibitem[{{Bouwens} {et~al.}(2012{\natexlab{a}}){Bouwens}, {Illingworth}, {Oesch}, {Franx}, {Labb{\'e}}, {Trenti}, {van Dokkum}, {Carollo}, {Gonz{\'a}lez}, {Smit}, \& {Magee}}]{Bouwens_2012a}
{Bouwens}, R.~J., {Illingworth}, G.~D., {Oesch}, P.~A., {et~al.} 2012{\natexlab{a}}, \apj, 754, 83, \dodoi{10.1088/0004-637X/754/2/83}

\bibitem[{{Bouwens} {et~al.}(2012{\natexlab{b}}){Bouwens}, {Illingworth}, {Oesch}, {Trenti}, {Labb{\'e}}, {Franx}, {Stiavelli}, {Carollo}, {van Dokkum}, \& {Magee}}]{Bouwens_2012b}
---. 2012{\natexlab{b}}, \apjl, 752, L5, \dodoi{10.1088/2041-8205/752/1/L5}

\bibitem[{{Brinchmann} {et~al.}(2004){Brinchmann}, {Charlot}, {White}, {Tremonti}, {Kauffmann}, {Heckman}, \& {Brinkmann}}]{Brinchmann_2004}
{Brinchmann}, J., {Charlot}, S., {White}, S.~D.~M., {et~al.} 2004, \mnras, 351, 1151, \dodoi{10.1111/j.1365-2966.2004.07881.x}

\bibitem[{{Bruzual} \& {Charlot}(2003)}]{Bruzual_2003}
{Bruzual}, G., \& {Charlot}, S. 2003, \mnras, 344, 1000, \dodoi{10.1046/j.1365-8711.2003.06897.x}

\bibitem[{{Calzetti} {et~al.}(2000){Calzetti}, {Armus}, {Bohlin}, {Kinney}, {Koornneef}, \& {Storchi-Bergmann}}]{Calzetti_2000}
{Calzetti}, D., {Armus}, L., {Bohlin}, R.~C., {et~al.} 2000, \apj, 533, 682, \dodoi{10.1086/308692}

\bibitem[{{Cappellari}(2012)}]{Cappellari_2012}
{Cappellari}, M. 2012, {pPXF: Penalized Pixel-Fitting stellar kinematics extraction}, Astrophysics Source Code Library, record ascl:1210.002

\bibitem[{{Caputi} {et~al.}(2011){Caputi}, {Cirasuolo}, {Dunlop}, {McLure}, {Farrah}, \& {Almaini}}]{Caputi_2011}
{Caputi}, K.~I., {Cirasuolo}, M., {Dunlop}, J.~S., {et~al.} 2011, \mnras, 413, 162, \dodoi{10.1111/j.1365-2966.2010.18118.x}

\bibitem[{{Caputi} {et~al.}(2015){Caputi}, {Ilbert}, {Laigle}, {McCracken}, {Le F{\`e}vre}, {Fynbo}, {Milvang-Jensen}, {Capak}, {Salvato}, \& {Taniguchi}}]{Caputi_2015}
{Caputi}, K.~I., {Ilbert}, O., {Laigle}, C., {et~al.} 2015, \apj, 810, 73, \dodoi{10.1088/0004-637X/810/1/73}

\bibitem[{{Carnall} {et~al.}(2018){Carnall}, {McLure}, {Dunlop}, \& {Dav{\'e}}}]{Carnall_2018}
{Carnall}, A.~C., {McLure}, R.~J., {Dunlop}, J.~S., \& {Dav{\'e}}, R. 2018, \mnras, 480, 4379, \dodoi{10.1093/mnras/sty2169}

\bibitem[{{Carnall} {et~al.}(2019){Carnall}, {McLure}, {Dunlop}, {Cullen}, {McLeod}, {Wild}, {Johnson}, {Appleby}, {Dav{\'e}}, {Amorin}, {Bolzonella}, {Castellano}, {Cimatti}, {Cucciati}, {Gargiulo}, {Garilli}, {Marchi}, {Pentericci}, {Pozzetti}, {Schreiber}, {Talia}, \& {Zamorani}}]{Carnall_2019a}
{Carnall}, A.~C., {McLure}, R.~J., {Dunlop}, J.~S., {et~al.} 2019, \mnras, 490, 417, \dodoi{10.1093/mnras/stz2544}

\bibitem[{{Chabrier}(2003)}]{Chabrier_2003}
{Chabrier}, G. 2003, \pasp, 115, 763, \dodoi{10.1086/376392}

\bibitem[{{Chauke} {et~al.}(2018){Chauke}, {van der Wel}, {Pacifici}, {Bezanson}, {Wu}, {Gallazzi}, {Noeske}, {Straatman}, {Mu{\~n}os-Mateos}, {Franx}, {Bari{\v{s}}i{\'c}}, {Bell}, {Brammer}, {Calhau}, {van Houdt}, {Labb{\'e}}, {Maseda}, {Muzzin}, {Rix}, \& {Sobral}}]{Chauke_2018}
{Chauke}, P., {van der Wel}, A., {Pacifici}, C., {et~al.} 2018, \apj, 861, 13, \dodoi{10.3847/1538-4357/aac324}

\bibitem[{{Chen} {et~al.}(2021){Chen}, {Mo}, {Li}, {Wang}, {Wang}, {Yang}, {Zhang}, \& {Katz}}]{Chen_2021}
{Chen}, Y., {Mo}, H.~J., {Li}, C., {et~al.} 2021, \mnras, 507, 2510, \dodoi{10.1093/mnras/stab2377}

\bibitem[{{Chen} \& {Wang}(2023)}]{Chen_2023}
{Chen}, Y., \& {Wang}, K. 2023, {HIPP: HIgh-Performance Package for scientific computation}, Astrophysics Source Code Library, record ascl:2301.030

\bibitem[{{Chen} {et~al.}(2012){Chen}, {Kauffmann}, {Tremonti}, {White}, {Heckman}, {Kova{\v{c}}}, {Bundy}, {Chisholm}, {Maraston}, {Schneider}, {Bolton}, {Weaver}, \& {Brinkmann}}]{Chen_2012}
{Chen}, Y.-M., {Kauffmann}, G., {Tremonti}, C.~A., {et~al.} 2012, \mnras, 421, 314, \dodoi{10.1111/j.1365-2966.2011.20306.x}

\bibitem[{{Chiosi}(1980)}]{Chiosi_1980}
{Chiosi}, C. 1980, \aap, 83, 206

\bibitem[{{Cid Fernandes} {et~al.}(2005){Cid Fernandes}, {Mateus}, {Sodr{\'e}}, {Stasi{\'n}ska}, \& {Gomes}}]{CidFernandes_2005}
{Cid Fernandes}, R., {Mateus}, A., {Sodr{\'e}}, L., {Stasi{\'n}ska}, G., \& {Gomes}, J.~M. 2005, \mnras, 358, 363, \dodoi{10.1111/j.1365-2966.2005.08752.x}

\bibitem[{{Collette} {et~al.}(2023){Collette}, {Kluyver}, {Caswell}, {Tocknell}, {Kieffer}, {Jelenak}, {Scopatz}, {Dale}, {Chen}, {VINCENT}, {Einhorn}, {Payno}, {Juliagarriga}, {Sciarelli}, {Valls}, {Ghosh}, {Kofoed Pedersen}, {Kittisopikul}, {Jakirkham}, {Raspaud}, {Danilevski}, {Abbasi}, {Readey}, {M{\"u}hlbauer}, {Paramonov}, {Chan}, {De Schepper}, {Sol{\'e}}, {Jialin}, \& {Hay Guest}}]{Collette_2023}
{Collette}, A., {Kluyver}, T., {Caswell}, T.~A., {et~al.} 2023, {h5py/h5py: 3.8.0-aarch64-wheels}, 3.8.0-aarch64-wheels,  Zenodo, \dodoi{10.5281/zenodo.7568214}

\bibitem[{{Conroy}(2013)}]{Conroy_2013}
{Conroy}, C. 2013, \araa, 51, 393, \dodoi{10.1146/annurev-astro-082812-141017}

\bibitem[{{Conroy} \& {Gunn}(2010)}]{Conroy_2010}
{Conroy}, C., \& {Gunn}, J.~E. 2010, \apj, 712, 833, \dodoi{10.1088/0004-637X/712/2/833}

\bibitem[{{Conroy} {et~al.}(2009){Conroy}, {Gunn}, \& {White}}]{Conroy_2009a}
{Conroy}, C., {Gunn}, J.~E., \& {White}, M. 2009, \apj, 699, 486, \dodoi{10.1088/0004-637X/699/1/486}

\bibitem[{{Conroy} \& {Wechsler}(2009)}]{Conroy_2009b}
{Conroy}, C., \& {Wechsler}, R.~H. 2009, \apj, 696, 620, \dodoi{10.1088/0004-637X/696/1/620}

\bibitem[{{Crain} \& {van de Voort}(2023)}]{Crain_2023}
{Crain}, R.~A., \& {van de Voort}, F. 2023, \araa, 61, 473, \dodoi{10.1146/annurev-astro-041923-043618}

\bibitem[{{Crain} {et~al.}(2009){Crain}, {Theuns}, {Dalla Vecchia}, {Eke}, {Frenk}, {Jenkins}, {Kay}, {Peacock}, {Pearce}, {Schaye}, {Springel}, {Thomas}, {White}, \& {Wiersma}}]{Crain_2009}
{Crain}, R.~A., {Theuns}, T., {Dalla Vecchia}, C., {et~al.} 2009, \mnras, 399, 1773, \dodoi{10.1111/j.1365-2966.2009.15402.x}

\bibitem[{{Cucciati} {et~al.}(2012){Cucciati}, {Tresse}, {Ilbert}, {Le F{\`e}vre}, {Garilli}, {Le Brun}, {Cassata}, {Franzetti}, {Maccagni}, {Scodeggio}, {Zucca}, {Zamorani}, {Bardelli}, {Bolzonella}, {Bielby}, {McCracken}, {Zanichelli}, \& {Vergani}}]{Cucciati_2012}
{Cucciati}, O., {Tresse}, L., {Ilbert}, O., {et~al.} 2012, \aap, 539, A31, \dodoi{10.1051/0004-6361/201118010}

\bibitem[{{Cui} {et~al.}(2024){Cui}, {Gu}, \& {Shi}}]{Cui_2024}
{Cui}, J., {Gu}, Q., \& {Shi}, Y. 2024, \mnras, 528, 2391, \dodoi{10.1093/mnras/stae156}

\bibitem[{{da Costa-Luis} {et~al.}(2024){da Costa-Luis}, {Larroque}, {Altendorf}, {Mary}, {richardsheridan}, {Korobov}, {Yorav-Raphael}, {Ivanov}, {Bargull}, {Rodrigues}, {Chen}, {Dektyarev}, {mjstevens777}, {Pagel}, {Zugnoni}, {JC}, {CrazyPython}, {Newey}, {Lee}, {pgajdos}, {Todd}, {Malmgren}, {redbug312}, {Desh}, {Nechaev}, {G{\'o}rny}, {Boyle}, {Nordlund}, {MapleCCC}, \& {McCracken}}]{daCosta-Luis_2024}
{da Costa-Luis}, C., {Larroque}, S.~K., {Altendorf}, K., {et~al.} 2024, {tqdm: A fast, Extensible Progress Bar for Python and CLI}, v4.66.5,  Zenodo, \dodoi{10.5281/zenodo.595120}

\bibitem[{{Dahlen} {et~al.}(2007){Dahlen}, {Mobasher}, {Dickinson}, {Ferguson}, {Giavalisco}, {Kretchmer}, \& {Ravindranath}}]{Dahlen_2007}
{Dahlen}, T., {Mobasher}, B., {Dickinson}, M., {et~al.} 2007, \apj, 654, 172, \dodoi{10.1086/508854}

\bibitem[{{Davis} {et~al.}(1985){Davis}, {Efstathiou}, {Frenk}, \& {White}}]{Davis_1985}
{Davis}, M., {Efstathiou}, G., {Frenk}, C.~S., \& {White}, S.~D.~M. 1985, \apj, 292, 371, \dodoi{10.1086/163168}

\bibitem[{{De Lucia} \& {Blaizot}(2007)}]{DeLucia_2007}
{De Lucia}, G., \& {Blaizot}, J. 2007, \mnras, 375, 2, \dodoi{10.1111/j.1365-2966.2006.11287.x}

\bibitem[{{Dekel} \& {Mandelker}(2014)}]{Dekel_2014}
{Dekel}, A., \& {Mandelker}, N. 2014, \mnras, 444, 2071, \dodoi{10.1093/mnras/stu1427}

\bibitem[{{Dekel} {et~al.}(2013){Dekel}, {Zolotov}, {Tweed}, {Cacciato}, {Ceverino}, \& {Primack}}]{Dekel_2013}
{Dekel}, A., {Zolotov}, A., {Tweed}, D., {et~al.} 2013, \mnras, 435, 999, \dodoi{10.1093/mnras/stt1338}

\bibitem[{{Diemer} {et~al.}(2017){Diemer}, {Sparre}, {Abramson}, \& {Torrey}}]{Diemer_2017}
{Diemer}, B., {Sparre}, M., {Abramson}, L.~E., \& {Torrey}, P. 2017, \apj, 839, 26, \dodoi{10.3847/1538-4357/aa68e5}

\bibitem[{{Dolag} {et~al.}(2009){Dolag}, {Borgani}, {Murante}, \& {Springel}}]{Dolag_2009}
{Dolag}, K., {Borgani}, S., {Murante}, G., \& {Springel}, V. 2009, \mnras, 399, 497, \dodoi{10.1111/j.1365-2966.2009.15034.x}

\bibitem[{{Donnari} {et~al.}(2021){Donnari}, {Pillepich}, {Nelson}, {Marinacci}, {Vogelsberger}, \& {Hernquist}}]{Donnari_2021}
{Donnari}, M., {Pillepich}, A., {Nelson}, D., {et~al.} 2021, \mnras, 506, 4760, \dodoi{10.1093/mnras/stab1950}

\bibitem[{{Donnari} {et~al.}(2019){Donnari}, {Pillepich}, {Nelson}, {Vogelsberger}, {Genel}, {Weinberger}, {Marinacci}, {Springel}, \& {Hernquist}}]{Donnari_2019}
---. 2019, \mnras, 485, 4817, \dodoi{10.1093/mnras/stz712}

\bibitem[{{Dou} {et~al.}(2021){Dou}, {Peng}, {Renzini}, {Ho}, {Mannucci}, {Daddi}, {Gao}, {Maiolino}, {Zhang}, {Gu}, {Li}, {Lilly}, {Pan}, {Yuan}, \& {Zheng}}]{Dou_2021}
{Dou}, J., {Peng}, Y., {Renzini}, A., {et~al.} 2021, \apj, 915, 94, \dodoi{10.3847/1538-4357/abfaf7}

\bibitem[{{Duncan} {et~al.}(2014){Duncan}, {Conselice}, {Mortlock}, {Hartley}, {Guo}, {Ferguson}, {Dav{\'e}}, {Lu}, {Ownsworth}, {Ashby}, {Dekel}, {Dickinson}, {Faber}, {Giavalisco}, {Grogin}, {Kocevski}, {Koekemoer}, {Somerville}, \& {White}}]{Duncan_2014}
{Duncan}, K., {Conselice}, C.~J., {Mortlock}, A., {et~al.} 2014, \mnras, 444, 2960, \dodoi{10.1093/mnras/stu1622}

\bibitem[{{Eales} {et~al.}(2017){Eales}, {de Vis}, {Smith}, {Appah}, {Ciesla}, {Duffield}, \& {Schofield}}]{Eales_2017}
{Eales}, S., {de Vis}, P., {Smith}, M. W.~L., {et~al.} 2017, \mnras, 465, 3125, \dodoi{10.1093/mnras/stw2875}

\bibitem[{{Faber}(1972)}]{Faber_1972}
{Faber}, S.~M. 1972, \aap, 20, 361

\bibitem[{{Feldmann}(2017)}]{Feldmann_2017}
{Feldmann}, R. 2017, \mnras, 470, L59, \dodoi{10.1093/mnrasl/slx073}

\bibitem[{{Finlator} {et~al.}(2007){Finlator}, {Dav{\'e}}, \& {Oppenheimer}}]{Finlator_2007}
{Finlator}, K., {Dav{\'e}}, R., \& {Oppenheimer}, B.~D. 2007, \mnras, 376, 1861, \dodoi{10.1111/j.1365-2966.2007.11578.x}

\bibitem[{{Gallazzi} {et~al.}(2008){Gallazzi}, {Brinchmann}, {Charlot}, \& {White}}]{Gallazzi_2008}
{Gallazzi}, A., {Brinchmann}, J., {Charlot}, S., \& {White}, S. D.~M. 2008, \mnras, 383, 1439, \dodoi{10.1111/j.1365-2966.2007.12632.x}

\bibitem[{{Gallazzi} {et~al.}(2005){Gallazzi}, {Charlot}, {Brinchmann}, {White}, \& {Tremonti}}]{Gallazzi_2005}
{Gallazzi}, A., {Charlot}, S., {Brinchmann}, J., {White}, S. D.~M., \& {Tremonti}, C.~A. 2005, \mnras, 362, 41, \dodoi{10.1111/j.1365-2966.2005.09321.x}

\bibitem[{{Gallazzi} {et~al.}(2021){Gallazzi}, {Pasquali}, {Zibetti}, \& {Barbera}}]{Gallazzi_2021}
{Gallazzi}, A.~R., {Pasquali}, A., {Zibetti}, S., \& {Barbera}, F.~L. 2021, \mnras, 502, 4457, \dodoi{10.1093/mnras/stab265}

\bibitem[{{Gladders} {et~al.}(2013){Gladders}, {Oemler}, {Dressler}, {Poggianti}, {Vulcani}, \& {Abramson}}]{Gladders_2013}
{Gladders}, M.~D., {Oemler}, A., {Dressler}, A., {et~al.} 2013, \apj, 770, 64, \dodoi{10.1088/0004-637X/770/1/64}

\bibitem[{{Gonz{\'a}lez} {et~al.}(2011){Gonz{\'a}lez}, {Labb{\'e}}, {Bouwens}, {Illingworth}, {Franx}, \& {Kriek}}]{Gonzalez_2011}
{Gonz{\'a}lez}, V., {Labb{\'e}}, I., {Bouwens}, R.~J., {et~al.} 2011, \apjl, 735, L34, \dodoi{10.1088/2041-8205/735/2/L34}

\bibitem[{{Grazian} {et~al.}(2015){Grazian}, {Fontana}, {Santini}, {Dunlop}, {Ferguson}, {Castellano}, {Amorin}, {Ashby}, {Barro}, {Behroozi}, {Boutsia}, {Caputi}, {Chary}, {Dekel}, {Dickinson}, {Faber}, {Fazio}, {Finkelstein}, {Galametz}, {Giallongo}, {Giavalisco}, {Grogin}, {Guo}, {Kocevski}, {Koekemoer}, {Koo}, {Lee}, {Lu}, {Merlin}, {Mobasher}, {Nonino}, {Papovich}, {Paris}, {Pentericci}, {Reddy}, {Renzini}, {Salmon}, {Salvato}, {Sommariva}, {Song}, \& {Vanzella}}]{Grazian_2015}
{Grazian}, A., {Fontana}, A., {Santini}, P., {et~al.} 2015, \aap, 575, A96, \dodoi{10.1051/0004-6361/201424750}

\bibitem[{{Gruppioni} {et~al.}(2013){Gruppioni}, {Pozzi}, {Rodighiero}, {Delvecchio}, {Berta}, {Pozzetti}, {Zamorani}, {Andreani}, {Cimatti}, {Ilbert}, {Le Floc'h}, {Lutz}, {Magnelli}, {Marchetti}, {Monaco}, {Nordon}, {Oliver}, {Popesso}, {Riguccini}, {Roseboom}, {Rosario}, {Sargent}, {Vaccari}, {Altieri}, {Aussel}, {Bongiovanni}, {Cepa}, {Daddi}, {Dom{\'\i}nguez-S{\'a}nchez}, {Elbaz}, {F{\"o}rster Schreiber}, {Genzel}, {Iribarrem}, {Magliocchetti}, {Maiolino}, {Poglitsch}, {P{\'e}rez Garc{\'\i}a}, {Sanchez-Portal}, {Sturm}, {Tacconi}, {Valtchanov}, {Amblard}, {Arumugam}, {Bethermin}, {Bock}, {Boselli}, {Buat}, {Burgarella}, {Castro-Rodr{\'\i}guez}, {Cava}, {Chanial}, {Clements}, {Conley}, {Cooray}, {Dowell}, {Dwek}, {Eales}, {Franceschini}, {Glenn}, {Griffin}, {Hatziminaoglou}, {Ibar}, {Isaak}, {Ivison}, {Lagache}, {Levenson}, {Lu}, {Madden}, {Maffei}, {Mainetti}, {Nguyen}, {O'Halloran}, {Page}, {Panuzzo}, {Papageorgiou}, {Pearson}, {P{\'e}rez-Fournon}, {Pohlen}, {Rigopoulou}, {Rowan-Robinson}, {Schulz},
  {Scott}, {Seymour}, {Shupe}, {Smith}, {Stevens}, {Symeonidis}, {Trichas}, {Tugwell}, {Vigroux}, {Wang}, {Wright}, {Xu}, {Zemcov}, {Bardelli}, {Carollo}, {Contini}, {Le F{\'e}vre}, {Lilly}, {Mainieri}, {Renzini}, {Scodeggio}, \& {Zucca}}]{Gruppioni_2013}
{Gruppioni}, C., {Pozzi}, F., {Rodighiero}, G., {et~al.} 2013, \mnras, 432, 23, \dodoi{10.1093/mnras/stt308}

\bibitem[{{Guo} {et~al.}(2011){Guo}, {White}, {Boylan-Kolchin}, {De Lucia}, {Kauffmann}, {Lemson}, {Li}, {Springel}, \& {Weinmann}}]{Guo_2011}
{Guo}, Q., {White}, S., {Boylan-Kolchin}, M., {et~al.} 2011, \mnras, 413, 101, \dodoi{10.1111/j.1365-2966.2010.18114.x}

\bibitem[{{Han} {et~al.}(2023){Han}, {Fan}, {Zheng}, {Bai}, \& {Han}}]{Han_2023}
{Han}, Y., {Fan}, L., {Zheng}, X.~Z., {Bai}, J.-M., \& {Han}, Z. 2023, \apjs, 269, 39, \dodoi{10.3847/1538-4365/acfc3a}

\bibitem[{{Harris} {et~al.}(2020){Harris}, {Millman}, {van der Walt}, {Gommers}, {Virtanen}, {Cournapeau}, {Wieser}, {Taylor}, {Berg}, {Smith}, {Kern}, {Picus}, {Hoyer}, {van Kerkwijk}, {Brett}, {Haldane}, {del R{\'\i}o}, {Wiebe}, {Peterson}, {G{\'e}rard-Marchant}, {Sheppard}, {Reddy}, {Weckesser}, {Abbasi}, {Gohlke}, \& {Oliphant}}]{Harris_2020}
{Harris}, C.~R., {Millman}, K.~J., {van der Walt}, S.~J., {et~al.} 2020, \nat, 585, 357, \dodoi{10.1038/s41586-020-2649-2}

\bibitem[{{Heavens} {et~al.}(2004){Heavens}, {Panter}, {Jimenez}, \& {Dunlop}}]{Heavens_2004}
{Heavens}, A., {Panter}, B., {Jimenez}, R., \& {Dunlop}, J. 2004, \nat, 428, 625, \dodoi{10.1038/nature02474}

\bibitem[{{Hunter}(2007)}]{Hunter_2007}
{Hunter}, J.~D. 2007, Computing in Science and Engineering, 9, 90, \dodoi{10.1109/MCSE.2007.55}

\bibitem[{{Ilbert} {et~al.}(2013){Ilbert}, {McCracken}, {Le F{\`e}vre}, {Capak}, {Dunlop}, {Karim}, {Renzini}, {Caputi}, {Boissier}, {Arnouts}, {Aussel}, {Comparat}, {Guo}, {Hudelot}, {Kartaltepe}, {Kneib}, {Krogager}, {Le Floc'h}, {Lilly}, {Mellier}, {Milvang-Jensen}, {Moutard}, {Onodera}, {Richard}, {Salvato}, {Sanders}, {Scoville}, {Silverman}, {Taniguchi}, {Tasca}, {Thomas}, {Toft}, {Tresse}, {Vergani}, {Wolk}, \& {Zirm}}]{Ilbert_2013}
{Ilbert}, O., {McCracken}, H.~J., {Le F{\`e}vre}, O., {et~al.} 2013, \aap, 556, A55, \dodoi{10.1051/0004-6361/201321100}

\bibitem[{{Ivezi{\'c}} {et~al.}(2019){Ivezi{\'c}}, {Kahn}, {Tyson}, {Abel}, {Acosta}, {Allsman}, {Alonso}, {AlSayyad}, {Anderson}, {Andrew}, \& et~al.}]{Ivezic_2019}
{Ivezi{\'c}}, {\v{Z}}., {Kahn}, S.~M., {Tyson}, J.~A., {et~al.} 2019, \apj, 873, 111, \dodoi{10.3847/1538-4357/ab042c}

\bibitem[{{Iyer} \& {Gawiser}(2017)}]{Iyer_2017}
{Iyer}, K., \& {Gawiser}, E. 2017, \apj, 838, 127, \dodoi{10.3847/1538-4357/aa63f0}

\bibitem[{{Iyer} {et~al.}(2019){Iyer}, {Gawiser}, {Faber}, {Ferguson}, {Kartaltepe}, {Koekemoer}, {Pacifici}, \& {Somerville}}]{Iyer_2019}
{Iyer}, K.~G., {Gawiser}, E., {Faber}, S.~M., {et~al.} 2019, \apj, 879, 116, \dodoi{10.3847/1538-4357/ab2052}

\bibitem[{{Jimenez} {et~al.}(2005){Jimenez}, {Panter}, {Heavens}, \& {Verde}}]{Jimenez_2005}
{Jimenez}, R., {Panter}, B., {Heavens}, A.~F., \& {Verde}, L. 2005, \mnras, 356, 495, \dodoi{10.1111/j.1365-2966.2004.08469.x}

\bibitem[{{Johnson} {et~al.}(2022){Johnson}, {Foreman-Mackey}, {Sick}, {Leja}, {Byler}, {Walmsley}, {Tollerud}, {Leung}, \& {Scott}}]{Johnson_2022}
{Johnson}, B., {Foreman-Mackey}, D., {Sick}, J., {et~al.} 2022, {dfm/python-fsps: python-fsps v0.4.2rc1}, v0.4.2rc1,  Zenodo, \dodoi{10.5281/zenodo.7113363}

\bibitem[{{Johnson} {et~al.}(2021){Johnson}, {Leja}, {Conroy}, \& {Speagle}}]{Johnson_2021}
{Johnson}, B.~D., {Leja}, J., {Conroy}, C., \& {Speagle}, J.~S. 2021, \apjs, 254, 22, \dodoi{10.3847/1538-4365/abef67}

\bibitem[{{Katz} {et~al.}(1992){Katz}, {Hernquist}, \& {Weinberg}}]{Katz_1992}
{Katz}, N., {Hernquist}, L., \& {Weinberg}, D.~H. 1992, \apjl, 399, L109, \dodoi{10.1086/186619}

\bibitem[{{Kauffmann} {et~al.}(1993){Kauffmann}, {White}, \& {Guiderdoni}}]{Kauffmann_1993}
{Kauffmann}, G., {White}, S.~D.~M., \& {Guiderdoni}, B. 1993, \mnras, 264, 201, \dodoi{10.1093/mnras/264.1.201}

\bibitem[{{Kauffmann} {et~al.}(2004){Kauffmann}, {White}, {Heckman}, {M{\'e}nard}, {Brinchmann}, {Charlot}, {Tremonti}, \& {Brinkmann}}]{Kauffmann_2004}
{Kauffmann}, G., {White}, S. D.~M., {Heckman}, T.~M., {et~al.} 2004, \mnras, 353, 713, \dodoi{10.1111/j.1365-2966.2004.08117.x}

\bibitem[{{Kauffmann} {et~al.}(2003){Kauffmann}, {Heckman}, {White}, {Charlot}, {Tremonti}, {Brinchmann}, {Bruzual}, {Peng}, {Seibert}, {Bernardi}, {Blanton}, {Brinkmann}, {Castander}, {Cs{\'a}bai}, {Fukugita}, {Ivezic}, {Munn}, {Nichol}, {Padmanabhan}, {Thakar}, {Weinberg}, \& {York}}]{Kauffmann_2003}
{Kauffmann}, G., {Heckman}, T.~M., {White}, S. D.~M., {et~al.} 2003, \mnras, 341, 33, \dodoi{10.1046/j.1365-8711.2003.06291.x}

\bibitem[{{Kikuchihara} {et~al.}(2020){Kikuchihara}, {Ouchi}, {Ono}, {Mawatari}, {Chevallard}, {Harikane}, {Kojima}, {Oguri}, {Bruzual}, \& {Charlot}}]{Kikuchihara_2020}
{Kikuchihara}, S., {Ouchi}, M., {Ono}, Y., {et~al.} 2020, \apj, 893, 60, \dodoi{10.3847/1538-4357/ab7dbe}

\bibitem[{{Kroupa}(2001)}]{Kroupa_2001}
{Kroupa}, P. 2001, \mnras, 322, 231, \dodoi{10.1046/j.1365-8711.2001.04022.x}

\bibitem[{{Lacey} \& {Fall}(1985)}]{Lacey_1985}
{Lacey}, C.~G., \& {Fall}, S.~M. 1985, \apj, 290, 154, \dodoi{10.1086/162970}

\bibitem[{{Lam} {et~al.}(2015){Lam}, {Pitrou}, \& {Seibert}}]{Lam_2015}
{Lam}, S.~K., {Pitrou}, A., \& {Seibert}, S. 2015, in Proc. Second Workshop on the LLVM Compiler Infrastructure in HPC, 1--6, \dodoi{10.1145/2833157.2833162}

\bibitem[{{Laureijs} {et~al.}(2011){Laureijs}, {Amiaux}, {Arduini}, {Augu{\`e}res}, {Brinchmann}, {Cole}, {Cropper}, {Dabin}, {Duvet}, {Ealet}, \& et~al.}]{Laureijs_2011}
{Laureijs}, R., {Amiaux}, J., {Arduini}, S., {et~al.} 2011, arXiv e-prints, arXiv:1110.3193, \dodoi{10.48550/arXiv.1110.3193}

\bibitem[{{Lee} {et~al.}(2010){Lee}, {Ferguson}, {Somerville}, {Wiklind}, \& {Giavalisco}}]{Lee_2010}
{Lee}, S.-K., {Ferguson}, H.~C., {Somerville}, R.~S., {Wiklind}, T., \& {Giavalisco}, M. 2010, \apj, 725, 1644, \dodoi{10.1088/0004-637X/725/2/1644}

\bibitem[{{Lee} {et~al.}(2009){Lee}, {Idzi}, {Ferguson}, {Somerville}, {Wiklind}, \& {Giavalisco}}]{Lee_2009}
{Lee}, S.-K., {Idzi}, R., {Ferguson}, H.~C., {et~al.} 2009, \apjs, 184, 100, \dodoi{10.1088/0067-0049/184/1/100}

\bibitem[{{Leja} {et~al.}(2019{\natexlab{a}}){Leja}, {Carnall}, {Johnson}, {Conroy}, \& {Speagle}}]{Leja_2019b}
{Leja}, J., {Carnall}, A.~C., {Johnson}, B.~D., {Conroy}, C., \& {Speagle}, J.~S. 2019{\natexlab{a}}, \apj, 876, 3, \dodoi{10.3847/1538-4357/ab133c}

\bibitem[{{Leja} {et~al.}(2017){Leja}, {Johnson}, {Conroy}, {van Dokkum}, \& {Byler}}]{Leja_2017}
{Leja}, J., {Johnson}, B.~D., {Conroy}, C., {van Dokkum}, P.~G., \& {Byler}, N. 2017, \apj, 837, 170, \dodoi{10.3847/1538-4357/aa5ffe}

\bibitem[{{Leja} {et~al.}(2019{\natexlab{b}}){Leja}, {Johnson}, {Conroy}, {van Dokkum}, {Speagle}, {Brammer}, {Momcheva}, {Skelton}, {Whitaker}, {Franx}, \& {Nelson}}]{Leja_2019a}
{Leja}, J., {Johnson}, B.~D., {Conroy}, C., {et~al.} 2019{\natexlab{b}}, \apj, 877, 140, \dodoi{10.3847/1538-4357/ab1d5a}

\bibitem[{{Li} {et~al.}(2023){Li}, {Ho}, {Shangguan}, {Zhuang}, \& {Li}}]{Li_2023}
{Li}, Y.~A., {Ho}, L.~C., {Shangguan}, J., {Zhuang}, M.-Y., \& {Li}, R. 2023, \apjs, 267, 17, \dodoi{10.3847/1538-4365/acd4b5}

\bibitem[{{Lilly} {et~al.}(2013){Lilly}, {Carollo}, {Pipino}, {Renzini}, \& {Peng}}]{Lilly_2013}
{Lilly}, S.~J., {Carollo}, C.~M., {Pipino}, A., {Renzini}, A., \& {Peng}, Y. 2013, \apj, 772, 119, \dodoi{10.1088/0004-637X/772/2/119}

\bibitem[{{Lilly} {et~al.}(1996){Lilly}, {Le Fevre}, {Hammer}, \& {Crampton}}]{Lilly_1996}
{Lilly}, S.~J., {Le Fevre}, O., {Hammer}, F., \& {Crampton}, D. 1996, \apjl, 460, L1, \dodoi{10.1086/309975}

\bibitem[{{Lu} {et~al.}(2015){Lu}, {Mo}, {Lu}, {Katz}, {Weinberg}, {van den Bosch}, \& {Yang}}]{Lu_2015}
{Lu}, Z., {Mo}, H.~J., {Lu}, Y., {et~al.} 2015, \mnras, 450, 1604, \dodoi{10.1093/mnras/stv667}

\bibitem[{{MacArthur}(2005)}]{MacArthur_2005}
{MacArthur}, L.~A. 2005, \apj, 623, 795, \dodoi{10.1086/428827}

\bibitem[{{Madau} \& {Dickinson}(2014)}]{Madau_2014}
{Madau}, P., \& {Dickinson}, M. 2014, \araa, 52, 415, \dodoi{10.1146/annurev-astro-081811-125615}

\bibitem[{{Madau} {et~al.}(1996){Madau}, {Ferguson}, {Dickinson}, {Giavalisco}, {Steidel}, \& {Fruchter}}]{Madau_1996a}
{Madau}, P., {Ferguson}, H.~C., {Dickinson}, M.~E., {et~al.} 1996, \mnras, 283, 1388, \dodoi{10.1093/mnras/283.4.1388}

\bibitem[{{Magnelli} {et~al.}(2011){Magnelli}, {Elbaz}, {Chary}, {Dickinson}, {Le Borgne}, {Frayer}, \& {Willmer}}]{Magnelli_2011}
{Magnelli}, B., {Elbaz}, D., {Chary}, R.~R., {et~al.} 2011, \aap, 528, A35, \dodoi{10.1051/0004-6361/200913941}

\bibitem[{{Magnelli} {et~al.}(2013){Magnelli}, {Popesso}, {Berta}, {Pozzi}, {Elbaz}, {Lutz}, {Dickinson}, {Altieri}, {Andreani}, {Aussel}, {B{\'e}thermin}, {Bongiovanni}, {Cepa}, {Charmandaris}, {Chary}, {Cimatti}, {Daddi}, {F{\"o}rster Schreiber}, {Genzel}, {Gruppioni}, {Harwit}, {Hwang}, {Ivison}, {Magdis}, {Maiolino}, {Murphy}, {Nordon}, {Pannella}, {P{\'e}rez Garc{\'\i}a}, {Poglitsch}, {Rosario}, {Sanchez-Portal}, {Santini}, {Scott}, {Sturm}, {Tacconi}, \& {Valtchanov}}]{Magnelli_2013}
{Magnelli}, B., {Popesso}, P., {Berta}, S., {et~al.} 2013, \aap, 553, A132, \dodoi{10.1051/0004-6361/201321371}

\bibitem[{{Maraston} {et~al.}(2010){Maraston}, {Pforr}, {Renzini}, {Daddi}, {Dickinson}, {Cimatti}, \& {Tonini}}]{Maraston_2010}
{Maraston}, C., {Pforr}, J., {Renzini}, A., {et~al.} 2010, \mnras, 407, 830, \dodoi{10.1111/j.1365-2966.2010.16973.x}

\bibitem[{{Marigo} \& {Girardi}(2007)}]{Marigo_2007}
{Marigo}, P., \& {Girardi}, L. 2007, \aap, 469, 239, \dodoi{10.1051/0004-6361:20066772}

\bibitem[{{Marigo} {et~al.}(2008){Marigo}, {Girardi}, {Bressan}, {Groenewegen}, {Silva}, \& {Granato}}]{Marigo_2008}
{Marigo}, P., {Girardi}, L., {Bressan}, A., {et~al.} 2008, \aap, 482, 883, \dodoi{10.1051/0004-6361:20078467}

\bibitem[{{Marinacci} {et~al.}(2018){Marinacci}, {Vogelsberger}, {Pakmor}, {Torrey}, {Springel}, {Hernquist}, {Nelson}, {Weinberger}, {Pillepich}, {Naiman}, \& {Genel}}]{Marinacci_2018}
{Marinacci}, F., {Vogelsberger}, M., {Pakmor}, R., {et~al.} 2018, \mnras, 480, 5113, \dodoi{10.1093/mnras/sty2206}

\bibitem[{{McLeod} {et~al.}(2021){McLeod}, {McLure}, {Dunlop}, {Cullen}, {Carnall}, \& {Duncan}}]{McLeod_2021}
{McLeod}, D.~J., {McLure}, R.~J., {Dunlop}, J.~S., {et~al.} 2021, \mnras, 503, 4413, \dodoi{10.1093/mnras/stab731}

\bibitem[{{Moster} {et~al.}(2010){Moster}, {Somerville}, {Maulbetsch}, {van den Bosch}, {Macci{\`o}}, {Naab}, \& {Oser}}]{Moster_2010}
{Moster}, B.~P., {Somerville}, R.~S., {Maulbetsch}, C., {et~al.} 2010, \apj, 710, 903, \dodoi{10.1088/0004-637X/710/2/903}

\bibitem[{{Muzzin} {et~al.}(2013){Muzzin}, {Marchesini}, {Stefanon}, {Franx}, {McCracken}, {Milvang-Jensen}, {Dunlop}, {Fynbo}, {Brammer}, {Labb{\'e}}, \& {van Dokkum}}]{Muzzin_2013}
{Muzzin}, A., {Marchesini}, D., {Stefanon}, M., {et~al.} 2013, \apj, 777, 18, \dodoi{10.1088/0004-637X/777/1/18}

\bibitem[{{Naiman} {et~al.}(2018){Naiman}, {Pillepich}, {Springel}, {Ramirez-Ruiz}, {Torrey}, {Vogelsberger}, {Pakmor}, {Nelson}, {Marinacci}, {Hernquist}, {Weinberger}, \& {Genel}}]{Naiman_2018}
{Naiman}, J.~P., {Pillepich}, A., {Springel}, V., {et~al.} 2018, \mnras, 477, 1206, \dodoi{10.1093/mnras/sty618}

\bibitem[{{Neistein} {et~al.}(2006){Neistein}, {van den Bosch}, \& {Dekel}}]{Neistein_2006}
{Neistein}, E., {van den Bosch}, F.~C., \& {Dekel}, A. 2006, \mnras, 372, 933, \dodoi{10.1111/j.1365-2966.2006.10918.x}

\bibitem[{{Nelan} {et~al.}(2005){Nelan}, {Smith}, {Hudson}, {Wegner}, {Lucey}, {Moore}, {Quinney}, \& {Suntzeff}}]{Nelan_2005}
{Nelan}, J.~E., {Smith}, R.~J., {Hudson}, M.~J., {et~al.} 2005, \apj, 632, 137, \dodoi{10.1086/431962}

\bibitem[{{Nelson} {et~al.}(2018){Nelson}, {Pillepich}, {Springel}, {Weinberger}, {Hernquist}, {Pakmor}, {Genel}, {Torrey}, {Vogelsberger}, {Kauffmann}, {Marinacci}, \& {Naiman}}]{Nelson_2018}
{Nelson}, D., {Pillepich}, A., {Springel}, V., {et~al.} 2018, \mnras, 475, 624, \dodoi{10.1093/mnras/stx3040}

\bibitem[{{Nelson} {et~al.}(2019){Nelson}, {Springel}, {Pillepich}, {Rodriguez-Gomez}, {Torrey}, {Genel}, {Vogelsberger}, {Pakmor}, {Marinacci}, {Weinberger}, {Kelley}, {Lovell}, {Diemer}, \& {Hernquist}}]{Nelson_2019}
{Nelson}, D., {Springel}, V., {Pillepich}, A., {et~al.} 2019, Computational Astrophysics and Cosmology, 6, 2, \dodoi{10.1186/s40668-019-0028-x}

\bibitem[{{Nersesian} {et~al.}(2024){Nersesian}, {van der Wel}, {Gallazzi}, {Leja}, {Bezanson}, {Bell}, {D'Eugenio}, {de Graaff}, {Kaushal}, {Martorano}, {Maseda}, \& {Zibetti}}]{Nersesian_2024}
{Nersesian}, A., {van der Wel}, A., {Gallazzi}, A., {et~al.} 2024, \aap, 681, A94, \dodoi{10.1051/0004-6361/202346769}

\bibitem[{{Noll} {et~al.}(2009){Noll}, {Burgarella}, {Giovannoli}, {Buat}, {Marcillac}, \& {Mu{\~n}oz-Mateos}}]{Noll_2009}
{Noll}, S., {Burgarella}, D., {Giovannoli}, E., {et~al.} 2009, \aap, 507, 1793, \dodoi{10.1051/0004-6361/200912497}

\bibitem[{{O'Connell}(1976)}]{O'Connell_1976}
{O'Connell}, R.~W. 1976, \apj, 206, 370, \dodoi{10.1086/154392}

\bibitem[{{Ocvirk} {et~al.}(2006){Ocvirk}, {Pichon}, {Lan{\c{c}}on}, \& {Thi{\'e}baut}}]{Ocvirk_2006}
{Ocvirk}, P., {Pichon}, C., {Lan{\c{c}}on}, A., \& {Thi{\'e}baut}, E. 2006, \mnras, 365, 46, \dodoi{10.1111/j.1365-2966.2005.09182.x}

\bibitem[{{Pacifici} {et~al.}(2012){Pacifici}, {Charlot}, {Blaizot}, \& {Brinchmann}}]{Pacifici_2012}
{Pacifici}, C., {Charlot}, S., {Blaizot}, J., \& {Brinchmann}, J. 2012, \mnras, 421, 2002, \dodoi{10.1111/j.1365-2966.2012.20431.x}

\bibitem[{{Pacifici} {et~al.}(2016){Pacifici}, {Kassin}, {Weiner}, {Holden}, {Gardner}, {Faber}, {Ferguson}, {Koo}, {Primack}, {Bell}, {Dekel}, {Gawiser}, {Giavalisco}, {Rafelski}, {Simons}, {Barro}, {Croton}, {Dav{\'e}}, {Fontana}, {Grogin}, {Koekemoer}, {Lee}, {Salmon}, {Somerville}, \& {Behroozi}}]{Pacifici_2016}
{Pacifici}, C., {Kassin}, S.~A., {Weiner}, B.~J., {et~al.} 2016, \apj, 832, 79, \dodoi{10.3847/0004-637X/832/1/79}

\bibitem[{{Papovich} {et~al.}(2001){Papovich}, {Dickinson}, \& {Ferguson}}]{Papovich_2001}
{Papovich}, C., {Dickinson}, M., \& {Ferguson}, H.~C. 2001, \apj, 559, 620, \dodoi{10.1086/322412}

\bibitem[{{Peng} {et~al.}(2012){Peng}, {Lilly}, {Renzini}, \& {Carollo}}]{Peng_2012}
{Peng}, Y.-j., {Lilly}, S.~J., {Renzini}, A., \& {Carollo}, M. 2012, \apj, 757, 4, \dodoi{10.1088/0004-637X/757/1/4}

\bibitem[{{Peng} \& {Maiolino}(2014)}]{Peng_2014}
{Peng}, Y.-j., \& {Maiolino}, R. 2014, \mnras, 443, 3643, \dodoi{10.1093/mnras/stu1288}

\bibitem[{{Peng} {et~al.}(2010){Peng}, {Lilly}, {Kova{\v{c}}}, {Bolzonella}, {Pozzetti}, {Renzini}, {Zamorani}, {Ilbert}, {Knobel}, {Iovino}, {Maier}, {Cucciati}, {Tasca}, {Carollo}, {Silverman}, {Kampczyk}, {de Ravel}, {Sanders}, {Scoville}, {Contini}, {Mainieri}, {Scodeggio}, {Kneib}, {Le F{\`e}vre}, {Bardelli}, {Bongiorno}, {Caputi}, {Coppa}, {de la Torre}, {Franzetti}, {Garilli}, {Lamareille}, {Le Borgne}, {Le Brun}, {Mignoli}, {Perez Montero}, {Pello}, {Ricciardelli}, {Tanaka}, {Tresse}, {Vergani}, {Welikala}, {Zucca}, {Oesch}, {Abbas}, {Barnes}, {Bordoloi}, {Bottini}, {Cappi}, {Cassata}, {Cimatti}, {Fumana}, {Hasinger}, {Koekemoer}, {Leauthaud}, {Maccagni}, {Marinoni}, {McCracken}, {Memeo}, {Meneux}, {Nair}, {Porciani}, {Presotto}, \& {Scaramella}}]{Peng_2010}
{Peng}, Y.-j., {Lilly}, S.~J., {Kova{\v{c}}}, K., {et~al.} 2010, \apj, 721, 193, \dodoi{10.1088/0004-637X/721/1/193}

\bibitem[{{Perez} \& {Granger}(2007)}]{Perez_2007}
{Perez}, F., \& {Granger}, B.~E. 2007, Computing in Science and Engineering, 9, 21, \dodoi{10.1109/MCSE.2007.53}

\bibitem[{{Pillepich} {et~al.}(2018{\natexlab{a}}){Pillepich}, {Nelson}, {Hernquist}, {Springel}, {Pakmor}, {Torrey}, {Weinberger}, {Genel}, {Naiman}, {Marinacci}, \& {Vogelsberger}}]{Pillepich_2018a}
{Pillepich}, A., {Nelson}, D., {Hernquist}, L., {et~al.} 2018{\natexlab{a}}, \mnras, 475, 648, \dodoi{10.1093/mnras/stx3112}

\bibitem[{{Pillepich} {et~al.}(2018{\natexlab{b}}){Pillepich}, {Springel}, {Nelson}, {Genel}, {Naiman}, {Pakmor}, {Hernquist}, {Torrey}, {Vogelsberger}, {Weinberger}, \& {Marinacci}}]{Pillepich_2018b}
{Pillepich}, A., {Springel}, V., {Nelson}, D., {et~al.} 2018{\natexlab{b}}, \mnras, 473, 4077, \dodoi{10.1093/mnras/stx2656}

\bibitem[{{Pilyugin} {et~al.}(2012){Pilyugin}, {Grebel}, \& {Mattsson}}]{Pilyugin_2012}
{Pilyugin}, L.~S., {Grebel}, E.~K., \& {Mattsson}, L. 2012, \mnras, 424, 2316, \dodoi{10.1111/j.1365-2966.2012.21398.x}

\bibitem[{{Planck Collaboration} {et~al.}(2016){Planck Collaboration}, {Ade}, {Aghanim}, {Arnaud}, {Ashdown}, {Aumont}, {Baccigalupi}, {Banday}, {Barreiro}, {Bartlett}, \& et~al.}]{PlanckCollaboration_2016}
{Planck Collaboration}, {Ade}, P.~A.~R., {Aghanim}, N., {et~al.} 2016, \aap, 594, A13, \dodoi{10.1051/0004-6361/201525830}

\bibitem[{{Reddy} \& {Steidel}(2009)}]{Reddy_2009}
{Reddy}, N.~A., \& {Steidel}, C.~C. 2009, \apj, 692, 778, \dodoi{10.1088/0004-637X/692/1/778}

\bibitem[{{Renzini}(2006)}]{Renzini_2006}
{Renzini}, A. 2006, \araa, 44, 141, \dodoi{10.1146/annurev.astro.44.051905.092450}

\bibitem[{{Renzini} \& {Peng}(2015)}]{Renzini_2015}
{Renzini}, A., \& {Peng}, Y.-j. 2015, \apjl, 801, L29, \dodoi{10.1088/2041-8205/801/2/L29}

\bibitem[{{Robotham} \& {Driver}(2011)}]{Robotham_2011}
{Robotham}, A.~S.~G., \& {Driver}, S.~P. 2011, \mnras, 413, 2570, \dodoi{10.1111/j.1365-2966.2011.18327.x}

\bibitem[{{Rodriguez-Gomez} {et~al.}(2015){Rodriguez-Gomez}, {Genel}, {Vogelsberger}, {Sijacki}, {Pillepich}, {Sales}, {Torrey}, {Snyder}, {Nelson}, {Springel}, {Ma}, \& {Hernquist}}]{Rodriguez-Gomez_2015}
{Rodriguez-Gomez}, V., {Genel}, S., {Vogelsberger}, M., {et~al.} 2015, \mnras, 449, 49, \dodoi{10.1093/mnras/stv264}

\bibitem[{{Salim} {et~al.}(2018){Salim}, {Boquien}, \& {Lee}}]{Salim_2018}
{Salim}, S., {Boquien}, M., \& {Lee}, J.~C. 2018, \apj, 859, 11, \dodoi{10.3847/1538-4357/aabf3c}

\bibitem[{{Salim} \& {Narayanan}(2020)}]{Salim_2020}
{Salim}, S., \& {Narayanan}, D. 2020, \araa, 58, 529, \dodoi{10.1146/annurev-astro-032620-021933}

\bibitem[{{Salim} {et~al.}(2007){Salim}, {Rich}, {Charlot}, {Brinchmann}, {Johnson}, {Schiminovich}, {Seibert}, {Mallery}, {Heckman}, {Forster}, {Friedman}, {Martin}, {Morrissey}, {Neff}, {Small}, {Wyder}, {Bianchi}, {Donas}, {Lee}, {Madore}, {Milliard}, {Szalay}, {Welsh}, \& {Yi}}]{Salim_2007}
{Salim}, S., {Rich}, R.~M., {Charlot}, S., {et~al.} 2007, \apjs, 173, 267, \dodoi{10.1086/519218}

\bibitem[{{Salim} {et~al.}(2016){Salim}, {Lee}, {Janowiecki}, {da Cunha}, {Dickinson}, {Boquien}, {Burgarella}, {Salzer}, \& {Charlot}}]{Salim_2016}
{Salim}, S., {Lee}, J.~C., {Janowiecki}, S., {et~al.} 2016, \apjs, 227, 2, \dodoi{10.3847/0067-0049/227/1/2}

\bibitem[{{S{\'a}nchez} {et~al.}(2012){S{\'a}nchez}, {Kennicutt}, {Gil de Paz}, {van de Ven}, {V{\'\i}lchez}, {Wisotzki}, {Walcher}, {Mast}, {Aguerri}, {Albiol-P{\'e}rez}, {Alonso-Herrero}, {Alves}, {Bakos}, {Bart{\'a}kov{\'a}}, {Bland-Hawthorn}, {Boselli}, {Bomans}, {Castillo-Morales}, {Cortijo-Ferrero}, {de Lorenzo-C{\'a}ceres}, {Del Olmo}, {Dettmar}, {D{\'\i}az}, {Ellis}, {Falc{\'o}n-Barroso}, {Flores}, {Gallazzi}, {Garc{\'\i}a-Lorenzo}, {Gonz{\'a}lez Delgado}, {Gruel}, {Haines}, {Hao}, {Husemann}, {Igl{\'e}sias-P{\'a}ramo}, {Jahnke}, {Johnson}, {Jungwiert}, {Kalinova}, {Kehrig}, {Kupko}, {L{\'o}pez-S{\'a}nchez}, {Lyubenova}, {Marino}, {M{\'a}rmol-Queralt{\'o}}, {M{\'a}rquez}, {Masegosa}, {Meidt}, {Mendez-Abreu}, {Monreal-Ibero}, {Montijo}, {Mour{\~a}o}, {Palacios-Navarro}, {Papaderos}, {Pasquali}, {Peletier}, {P{\'e}rez}, {P{\'e}rez}, {Quirrenbach}, {Rela{\~n}o}, {Rosales-Ortega}, {Roth}, {Ruiz-Lara}, {S{\'a}nchez-Bl{\'a}zquez}, {Sengupta}, {Singh}, {Stanishev}, {Trager}, {Vazdekis}, {Viironen}, {Wild},
  {Zibetti}, \& {Ziegler}}]{Sanchez_2012}
{S{\'a}nchez}, S.~F., {Kennicutt}, R.~C., {Gil de Paz}, A., {et~al.} 2012, \aap, 538, A8, \dodoi{10.1051/0004-6361/201117353}

\bibitem[{{S{\'a}nchez} {et~al.}(2019){S{\'a}nchez}, {Avila-Reese}, {Rodr{\'\i}guez-Puebla}, {Ibarra-Medel}, {Calette}, {Bershady}, {Hern{\'a}ndez-Toledo}, {Pan}, \& {Bizyaev}}]{Sanchez_2019}
{S{\'a}nchez}, S.~F., {Avila-Reese}, V., {Rodr{\'\i}guez-Puebla}, A., {et~al.} 2019, \mnras, 482, 1557, \dodoi{10.1093/mnras/sty2730}

\bibitem[{{S{\'a}nchez-Bl{\'a}zquez} {et~al.}(2006){S{\'a}nchez-Bl{\'a}zquez}, {Peletier}, {Jim{\'e}nez-Vicente}, {Cardiel}, {Cenarro}, {Falc{\'o}n-Barroso}, {Gorgas}, {Selam}, \& {Vazdekis}}]{Sanchez-Blazquez_2006}
{S{\'a}nchez-Bl{\'a}zquez}, P., {Peletier}, R.~F., {Jim{\'e}nez-Vicente}, J., {et~al.} 2006, \mnras, 371, 703, \dodoi{10.1111/j.1365-2966.2006.10699.x}

\bibitem[{{Sanders} {et~al.}(2003){Sanders}, {Mazzarella}, {Kim}, {Surace}, \& {Soifer}}]{Sanders_2003}
{Sanders}, D.~B., {Mazzarella}, J.~M., {Kim}, D.~C., {Surace}, J.~A., \& {Soifer}, B.~T. 2003, \aj, 126, 1607, \dodoi{10.1086/376841}

\bibitem[{{Santini} {et~al.}(2012){Santini}, {Fontana}, {Grazian}, {Salimbeni}, {Fontanot}, {Paris}, {Boutsia}, {Castellano}, {Fiore}, {Gallozzi}, {Giallongo}, {Koekemoer}, {Menci}, {Pentericci}, \& {Somerville}}]{Santini_2012}
{Santini}, P., {Fontana}, A., {Grazian}, A., {et~al.} 2012, \aap, 538, A33, \dodoi{10.1051/0004-6361/201117513}

\bibitem[{{Schaye} {et~al.}(2015){Schaye}, {Crain}, {Bower}, {Furlong}, {Schaller}, {Theuns}, {Dalla Vecchia}, {Frenk}, {McCarthy}, {Helly}, {Jenkins}, {Rosas-Guevara}, {White}, {Baes}, {Booth}, {Camps}, {Navarro}, {Qu}, {Rahmati}, {Sawala}, {Thomas}, \& {Trayford}}]{Schaye_2015}
{Schaye}, J., {Crain}, R.~A., {Bower}, R.~G., {et~al.} 2015, \mnras, 446, 521, \dodoi{10.1093/mnras/stu2058}

\bibitem[{{Schaye} {et~al.}(2023){Schaye}, {Kugel}, {Schaller}, {Helly}, {Braspenning}, {Elbers}, {McCarthy}, {van Daalen}, {Vandenbroucke}, {Frenk}, {Kwan}, {Salcido}, {Bah{\'e}}, {Borrow}, {Chaikin}, {Hahn}, {Hu{\v{s}}ko}, {Jenkins}, {Lacey}, \& {Nobels}}]{Schaye_2023}
{Schaye}, J., {Kugel}, R., {Schaller}, M., {et~al.} 2023, \mnras, 526, 4978, \dodoi{10.1093/mnras/stad2419}

\bibitem[{{Schenker} {et~al.}(2013){Schenker}, {Robertson}, {Ellis}, {Ono}, {McLure}, {Dunlop}, {Koekemoer}, {Bowler}, {Ouchi}, {Curtis-Lake}, {Rogers}, {Schneider}, {Charlot}, {Stark}, {Furlanetto}, \& {Cirasuolo}}]{Schenker_2013}
{Schenker}, M.~A., {Robertson}, B.~E., {Ellis}, R.~S., {et~al.} 2013, \apj, 768, 196, \dodoi{10.1088/0004-637X/768/2/196}

\bibitem[{{Schiminovich} {et~al.}(2005){Schiminovich}, {Ilbert}, {Arnouts}, {Milliard}, {Tresse}, {Le F{\`e}vre}, {Treyer}, {Wyder}, {Budav{\'a}ri}, {Zucca}, {Zamorani}, {Martin}, {Adami}, {Arnaboldi}, {Bardelli}, {Barlow}, {Bianchi}, {Bolzonella}, {Bottini}, {Byun}, {Cappi}, {Contini}, {Charlot}, {Donas}, {Forster}, {Foucaud}, {Franzetti}, {Friedman}, {Garilli}, {Gavignaud}, {Guzzo}, {Heckman}, {Hoopes}, {Iovino}, {Jelinsky}, {Le Brun}, {Lee}, {Maccagni}, {Madore}, {Malina}, {Marano}, {Marinoni}, {McCracken}, {Mazure}, {Meneux}, {Morrissey}, {Neff}, {Paltani}, {Pell{\`o}}, {Picat}, {Pollo}, {Pozzetti}, {Radovich}, {Rich}, {Scaramella}, {Scodeggio}, {Seibert}, {Siegmund}, {Small}, {Szalay}, {Vettolani}, {Welsh}, {Xu}, \& {Zanichelli}}]{Schiminovich_2005}
{Schiminovich}, D., {Ilbert}, O., {Arnouts}, S., {et~al.} 2005, \apjl, 619, L47, \dodoi{10.1086/427077}

\bibitem[{{Schmidt}(1959)}]{Schmidt_1959}
{Schmidt}, M. 1959, \apj, 129, 243, \dodoi{10.1086/146614}

\bibitem[{{Schmidt}(1968)}]{Schmidt_1968}
---. 1968, \apj, 151, 393, \dodoi{10.1086/149446}

\bibitem[{{Shamshiri} {et~al.}(2015){Shamshiri}, {Thomas}, {Henriques}, {Tojeiro}, {Lemson}, {Oliver}, \& {Wilkins}}]{Shamshiri_2015}
{Shamshiri}, S., {Thomas}, P.~A., {Henriques}, B.~M., {et~al.} 2015, \mnras, 451, 2681, \dodoi{10.1093/mnras/stv883}

\bibitem[{{Simha} {et~al.}(2014){Simha}, {Weinberg}, {Conroy}, {Dave}, {Fardal}, {Katz}, \& {Oppenheimer}}]{Simha_2014}
{Simha}, V., {Weinberg}, D.~H., {Conroy}, C., {et~al.} 2014, arXiv e-prints, arXiv:1404.0402, \dodoi{10.48550/arXiv.1404.0402}

\bibitem[{{Somerville} \& {Primack}(1999)}]{Somerville_1999}
{Somerville}, R.~S., \& {Primack}, J.~R. 1999, \mnras, 310, 1087, \dodoi{10.1046/j.1365-8711.1999.03032.x}

\bibitem[{{Springel}(2010)}]{Springel_2010}
{Springel}, V. 2010, \mnras, 401, 791, \dodoi{10.1111/j.1365-2966.2009.15715.x}

\bibitem[{{Springel} {et~al.}(2001){Springel}, {White}, {Tormen}, \& {Kauffmann}}]{Springel_2001}
{Springel}, V., {White}, S. D.~M., {Tormen}, G., \& {Kauffmann}, G. 2001, \mnras, 328, 726, \dodoi{10.1046/j.1365-8711.2001.04912.x}

\bibitem[{{Springel} {et~al.}(2018){Springel}, {Pakmor}, {Pillepich}, {Weinberger}, {Nelson}, {Hernquist}, {Vogelsberger}, {Genel}, {Torrey}, {Marinacci}, \& {Naiman}}]{Springel_2018}
{Springel}, V., {Pakmor}, R., {Pillepich}, A., {et~al.} 2018, \mnras, 475, 676, \dodoi{10.1093/mnras/stx3304}

\bibitem[{{Takeuchi} {et~al.}(2003){Takeuchi}, {Yoshikawa}, \& {Ishii}}]{Takeuchi_2003}
{Takeuchi}, T.~T., {Yoshikawa}, K., \& {Ishii}, T.~T. 2003, \apjl, 587, L89, \dodoi{10.1086/375181}

\bibitem[{{Talbot} \& {Arnett}(1971)}]{Talbot_1971}
{Talbot}, Raymond~J., J., \& {Arnett}, W.~D. 1971, \apj, 170, 409, \dodoi{10.1086/151228}

\bibitem[{{Terrazas} {et~al.}(2020){Terrazas}, {Bell}, {Pillepich}, {Nelson}, {Somerville}, {Genel}, {Weinberger}, {Habouzit}, {Li}, {Hernquist}, \& {Vogelsberger}}]{Terrazas_2020}
{Terrazas}, B.~A., {Bell}, E.~F., {Pillepich}, A., {et~al.} 2020, \mnras, 493, 1888, \dodoi{10.1093/mnras/staa374}

\bibitem[{{The pandas development Team}(2024)}]{ThepandasdevelopmentTeam_2024}
{The pandas development Team}. 2024, {pandas-dev/pandas: Pandas}, v2.2.2,  Zenodo, \dodoi{10.5281/zenodo.3509134}

\bibitem[{{Thomas} {et~al.}(2005){Thomas}, {Maraston}, {Bender}, \& {Mendes de Oliveira}}]{Thomas_2005}
{Thomas}, D., {Maraston}, C., {Bender}, R., \& {Mendes de Oliveira}, C. 2005, \apj, 621, 673, \dodoi{10.1086/426932}

\bibitem[{{Thorne} {et~al.}(2021){Thorne}, {Robotham}, {Davies}, {Bellstedt}, {Driver}, {Bravo}, {Bremer}, {Holwerda}, {Hopkins}, {Lagos}, {Phillipps}, {Siudek}, {Taylor}, \& {Wright}}]{Thorne_2021}
{Thorne}, J.~E., {Robotham}, A. S.~G., {Davies}, L. J.~M., {et~al.} 2021, \mnras, 505, 540, \dodoi{10.1093/mnras/stab1294}

\bibitem[{{Tinsley}(1974)}]{Tinsley_1974}
{Tinsley}, B.~M. 1974, \apj, 192, 629, \dodoi{10.1086/153099}

\bibitem[{{Tinsley}(1980)}]{Tinsley_1980}
---. 1980, \fcp, 5, 287, \dodoi{10.48550/arXiv.2203.02041}

\bibitem[{{Trussler} {et~al.}(2020){Trussler}, {Maiolino}, {Maraston}, {Peng}, {Thomas}, {Goddard}, \& {Lian}}]{Trussler_2020}
{Trussler}, J., {Maiolino}, R., {Maraston}, C., {et~al.} 2020, \mnras, 491, 5406, \dodoi{10.1093/mnras/stz3286}

\bibitem[{{van der Wel} {et~al.}(2016){van der Wel}, {Noeske}, {Bezanson}, {Pacifici}, {Gallazzi}, {Franx}, {Mu{\~n}oz-Mateos}, {Bell}, {Brammer}, {Charlot}, {Chauk{\'e}}, {Labb{\'e}}, {Maseda}, {Muzzin}, {Rix}, {Sobral}, {van de Sande}, {van Dokkum}, {Wild}, \& {Wolf}}]{vanderWel_2016}
{van der Wel}, A., {Noeske}, K., {Bezanson}, R., {et~al.} 2016, \apjs, 223, 29, \dodoi{10.3847/0067-0049/223/2/29}

\bibitem[{{van der Wel} {et~al.}(2021){van der Wel}, {Bezanson}, {D'Eugenio}, {Straatman}, {Franx}, {van Houdt}, {Maseda}, {Gallazzi}, {Wu}, {Pacifici}, {Barisic}, {Brammer}, {Munoz-Mateos}, {Vervalcke}, {Zibetti}, {Sobral}, {de Graaff}, {Calhau}, {Kaushal}, {Muzzin}, {Bell}, \& {van Dokkum}}]{vanderWel_2021}
{van der Wel}, A., {Bezanson}, R., {D'Eugenio}, F., {et~al.} 2021, \apjs, 256, 44, \dodoi{10.3847/1538-4365/ac1356}

\bibitem[{{Virtanen} {et~al.}(2020){Virtanen}, {Gommers}, {Oliphant}, {Haberland}, {Reddy}, {Cournapeau}, {Burovski}, {Peterson}, {Weckesser}, {Bright}, {van der Walt}, {Brett}, {Wilson}, {Millman}, {Mayorov}, {Nelson}, {Jones}, {Kern}, {Larson}, {Carey}, {Polat}, {Feng}, {Moore}, {VanderPlas}, {Laxalde}, {Perktold}, {Cimrman}, {Henriksen}, {Quintero}, {Harris}, {Archibald}, {Ribeiro}, {Pedregosa}, {van Mulbregt}, \& {SciPy 1. 0 Contributors}}]{Virtanen_2020}
{Virtanen}, P., {Gommers}, R., {Oliphant}, T.~E., {et~al.} 2020, Nature Methods, 17, 261, \dodoi{10.1038/s41592-019-0686-2}

\bibitem[{{Walcher} {et~al.}(2015){Walcher}, {Coelho}, {Gallazzi}, {Bruzual}, {Charlot}, \& {Chiappini}}]{Walcher_2015}
{Walcher}, C.~J., {Coelho}, P.~R.~T., {Gallazzi}, A., {et~al.} 2015, \aap, 582, A46, \dodoi{10.1051/0004-6361/201525924}

\bibitem[{{Wang} \& {Lilly}(2021)}]{Wang_2021}
{Wang}, E., \& {Lilly}, S.~J. 2021, \apj, 910, 137, \dodoi{10.3847/1538-4357/abe413}

\bibitem[{{Wang} \& {Lilly}(2022)}]{Wang_2022}
---. 2022, \apj, 929, 95, \dodoi{10.3847/1538-4357/ac5e31}

\bibitem[{{Wang} {et~al.}(2023){Wang}, {Peng}, \& {Chen}}]{Wang_2023}
{Wang}, K., {Peng}, Y., \& {Chen}, Y. 2023, \mnras, 523, 1268, \dodoi{10.1093/mnras/stad1169}

\bibitem[{{Waskom}(2021)}]{Waskom_2021}
{Waskom}, M. 2021, The Journal of Open Source Software, 6, 3021, \dodoi{10.21105/joss.03021}

\bibitem[{{Weaver} {et~al.}(2023){Weaver}, {Davidzon}, {Toft}, {Ilbert}, {McCracken}, {Gould}, {Jespersen}, {Steinhardt}, {Lagos}, {Capak}, {Casey}, {Chartab}, {Faisst}, {Hayward}, {Kartaltepe}, {Kauffmann}, {Koekemoer}, {Kokorev}, {Laigle}, {Liu}, {Long}, {Magdis}, {McPartland}, {Milvang-Jensen}, {Mobasher}, {Moneti}, {Peng}, {Sanders}, {Shuntov}, {Sneppen}, {Valentino}, {Zalesky}, \& {Zamorani}}]{Weaver_2023}
{Weaver}, J.~R., {Davidzon}, I., {Toft}, S., {et~al.} 2023, \aap, 677, A184, \dodoi{10.1051/0004-6361/202245581}

\bibitem[{{Wetzel} {et~al.}(2013){Wetzel}, {Tinker}, {Conroy}, \& {van den Bosch}}]{Wetzel_2013}
{Wetzel}, A.~R., {Tinker}, J.~L., {Conroy}, C., \& {van den Bosch}, F.~C. 2013, \mnras, 432, 336, \dodoi{10.1093/mnras/stt469}

\bibitem[{{White} \& {Frenk}(1991)}]{White_1991}
{White}, S. D.~M., \& {Frenk}, C.~S. 1991, \apj, 379, 52, \dodoi{10.1086/170483}

\bibitem[{{Worthey} {et~al.}(1994){Worthey}, {Faber}, {Gonzalez}, \& {Burstein}}]{Worthey_1994}
{Worthey}, G., {Faber}, S.~M., {Gonzalez}, J.~J., \& {Burstein}, D. 1994, \apjs, 94, 687, \dodoi{10.1086/192087}

\bibitem[{{Worthey} \& {Ottaviani}(1997)}]{Worthey_1997}
{Worthey}, G., \& {Ottaviani}, D.~L. 1997, \apjs, 111, 377, \dodoi{10.1086/313021}

\bibitem[{{Wright} {et~al.}(2018){Wright}, {Driver}, \& {Robotham}}]{Wright_2018}
{Wright}, A.~H., {Driver}, S.~P., \& {Robotham}, A.~S.~G. 2018, \mnras, 480, 3491, \dodoi{10.1093/mnras/sty2136}

\bibitem[{{Wu} {et~al.}(2018){Wu}, {van der Wel}, {Gallazzi}, {Bezanson}, {Pacifici}, {Straatman}, {Franx}, {Bari{\v{s}}i{\'c}}, {Bell}, {Brammer}, {Calhau}, {Chauke}, {van Houdt}, {Maseda}, {Muzzin}, {Rix}, {Sobral}, {Spilker}, {van de Sande}, {van Dokkum}, \& {Wild}}]{Wu_2018}
{Wu}, P.-F., {van der Wel}, A., {Gallazzi}, A., {et~al.} 2018, \apj, 855, 85, \dodoi{10.3847/1538-4357/aab0a6}

\bibitem[{{Wu} {et~al.}(2021){Wu}, {Nelson}, {van der Wel}, {Pillepich}, {Zibetti}, {Bezanson}, {DEugenio}, {Gallazzi}, {Pacifici}, {Straatman}, {Bari{\v{s}}i{\'c}}, {Bell}, {Maseda}, {Muzzin}, {Sobral}, \& {Whitaker}}]{Wu_2021}
{Wu}, P.-F., {Nelson}, D., {van der Wel}, A., {et~al.} 2021, \aj, 162, 201, \dodoi{10.3847/1538-3881/ac20d6}

\bibitem[{{Wyder} {et~al.}(2005){Wyder}, {Treyer}, {Milliard}, {Schiminovich}, {Arnouts}, {Budav{\'a}ri}, {Barlow}, {Bianchi}, {Byun}, {Donas}, {Forster}, {Friedman}, {Heckman}, {Jelinsky}, {Lee}, {Madore}, {Malina}, {Martin}, {Morrissey}, {Neff}, {Rich}, {Siegmund}, {Small}, {Szalay}, \& {Welsh}}]{Wyder_2005}
{Wyder}, T.~K., {Treyer}, M.~A., {Milliard}, B., {et~al.} 2005, \apjl, 619, L15, \dodoi{10.1086/424735}

\bibitem[{{Yang} {et~al.}(2007){Yang}, {Mo}, {van den Bosch}, {Pasquali}, {Li}, \& {Barden}}]{Yang_2007}
{Yang}, X., {Mo}, H.~J., {van den Bosch}, F.~C., {et~al.} 2007, \apj, 671, 153, \dodoi{10.1086/522027}

\bibitem[{{Yang} {et~al.}(2012){Yang}, {Mo}, {van den Bosch}, {Zhang}, \& {Han}}]{Yang_2012}
{Yang}, X., {Mo}, H.~J., {van den Bosch}, F.~C., {Zhang}, Y., \& {Han}, J. 2012, \apj, 752, 41, \dodoi{10.1088/0004-637X/752/1/41}

\bibitem[{{York} {et~al.}(2000){York}, {Adelman}, {Anderson}, {Anderson}, {Annis}, {Bahcall}, {Bakken}, {Barkhouser}, {Bastian}, {Berman}, {Boroski}, {Bracker}, {Briegel}, {Briggs}, {Brinkmann}, {Brunner}, {Burles}, {Carey}, {Carr}, {Castander}, {Chen}, {Colestock}, {Connolly}, {Crocker}, {Csabai}, {Czarapata}, {Davis}, {Doi}, {Dombeck}, {Eisenstein}, {Ellman}, {Elms}, {Evans}, {Fan}, {Federwitz}, {Fiscelli}, {Friedman}, {Frieman}, {Fukugita}, {Gillespie}, {Gunn}, {Gurbani}, {de Haas}, {Haldeman}, {Harris}, {Hayes}, {Heckman}, {Hennessy}, {Hindsley}, {Holm}, {Holmgren}, {Huang}, {Hull}, {Husby}, {Ichikawa}, {Ichikawa}, {Ivezi{\'c}}, {Kent}, {Kim}, {Kinney}, {Klaene}, {Kleinman}, {Kleinman}, {Knapp}, {Korienek}, {Kron}, {Kunszt}, {Lamb}, {Lee}, {Leger}, {Limmongkol}, {Lindenmeyer}, {Long}, {Loomis}, {Loveday}, {Lucinio}, {Lupton}, {MacKinnon}, {Mannery}, {Mantsch}, {Margon}, {McGehee}, {McKay}, {Meiksin}, {Merelli}, {Monet}, {Munn}, {Narayanan}, {Nash}, {Neilsen}, {Neswold}, {Newberg}, {Nichol}, {Nicinski},
  {Nonino}, {Okada}, {Okamura}, {Ostriker}, {Owen}, {Pauls}, {Peoples}, {Peterson}, {Petravick}, {Pier}, {Pope}, {Pordes}, {Prosapio}, {Rechenmacher}, {Quinn}, {Richards}, {Richmond}, {Rivetta}, {Rockosi}, {Ruthmansdorfer}, {Sandford}, {Schlegel}, {Schneider}, {Sekiguchi}, {Sergey}, {Shimasaku}, {Siegmund}, {Smee}, {Smith}, {Snedden}, {Stone}, {Stoughton}, {Strauss}, {Stubbs}, {SubbaRao}, {Szalay}, {Szapudi}, {Szokoly}, {Thakar}, {Tremonti}, {Tucker}, {Uomoto}, {Vanden Berk}, {Vogeley}, {Waddell}, {Wang}, {Watanabe}, {Weinberg}, {Yanny}, {Yasuda}, \& {SDSS Collaboration}}]{York_2000}
{York}, D.~G., {Adelman}, J., {Anderson}, John~E., J., {et~al.} 2000, \aj, 120, 1579, \dodoi{10.1086/301513}

\bibitem[{{Zhan}(2011)}]{Zhan_2011}
{Zhan}, H. 2011, Scientia Sinica Physica, Mechanica \& Astronomica, 41, 1441, \dodoi{10.1360/132011-961}

\bibitem[{{Zhou} {et~al.}(2022){Zhou}, {Merrifield}, \& {Arag{\'o}n-Salamanca}}]{Zhou_2022}
{Zhou}, S., {Merrifield}, M., \& {Arag{\'o}n-Salamanca}, A. 2022, \mnras, 513, 5446, \dodoi{10.1093/mnras/stac1279}

\bibitem[{{Zhou} {et~al.}(2020){Zhou}, {Mo}, {Li}, {Boquien}, \& {Rossi}}]{Zhou_2020}
{Zhou}, S., {Mo}, H.~J., {Li}, C., {Boquien}, M., \& {Rossi}, G. 2020, \mnras, 497, 4753, \dodoi{10.1093/mnras/staa2337}

\bibitem[{{Zibetti} {et~al.}(2020){Zibetti}, {Gallazzi}, {Hirschmann}, {Consolandi}, {Falc{\'o}n-Barroso}, {van de Ven}, \& {Lyubenova}}]{Zibetti_2020}
{Zibetti}, S., {Gallazzi}, A.~R., {Hirschmann}, M., {et~al.} 2020, \mnras, 491, 3562, \dodoi{10.1093/mnras/stz3205}

\bibitem[{{Zibetti} {et~al.}(2017){Zibetti}, {Gallazzi}, {Ascasibar}, {Charlot}, {Galbany}, {Garc{\'\i}a Benito}, {Kehrig}, {de Lorenzo-C{\'a}ceres}, {Lyubenova}, {Marino}, {M{\'a}rquez}, {S{\'a}nchez}, {van de Ven}, {Walcher}, \& {Wisotzki}}]{Zibetti_2017}
{Zibetti}, S., {Gallazzi}, A.~R., {Ascasibar}, Y., {et~al.} 2017, \mnras, 468, 1902, \dodoi{10.1093/mnras/stx251}

\end{thebibliography}
\bibliographystyle{aasjournal}
\appendix
\section{Parameterization of SFH}
\label{app:sfh_binning_test}

This Appendix discusses our method for parameterizing SFHs of galaxies within the IllustrisTNG framework. Instead of relying solely on discrete snapshots, our approach integrates the formation times and initial masses of stellar particles over continuous periods to accurately reflect the galaxy's current spectral energy distribution. This method accounts for the dynamic changes in stellar particle locations due to processes such as mergers and interactions.

We adopt a refined SFH binning algorithm adapted from \citet{Shamshiri_2015}, where we set the nearest bin to the present epoch ($z=0$) at $t_0 = 0.01$ Gyr. The bins increase exponentially back in time:

\begin{equation}
    b^0t_0,\, b^0t_0,\, b^1t_0,\, b^1t_0,\, ...,\, b^nt_0,\, b^nt_0
\end{equation}

where $b$ is the scaling factor and $2(n+1)$ the total number of bins. The age of the universe, $T$, is thus:

\begin{equation}
    T = \sum_{i=0}^{n} 2b^it_0 = 2\frac{b^{n+1}-1}{b-1}t_0
\end{equation}

Determining the values for $t_0$, $2(n+1)$, and $b$ allows us to define the time intervals for each bin accurately. 
We conduct a test to evaluate the efficacy of this algorithm in recovering the actual spectra of galaxies and to determine the optimal number of bins required for reconstruction. For this test, we perform a stellar population analysis on galaxies in the TNG100 simulation using \textsc{fsps} and compare the spectra generated with different binning algorithms and bin numbers. Since our objective is to test the SFH, we only consider the continuum from the stellar component.

\begin{figure*}
    \centering
    \includegraphics[width=0.7\linewidth]{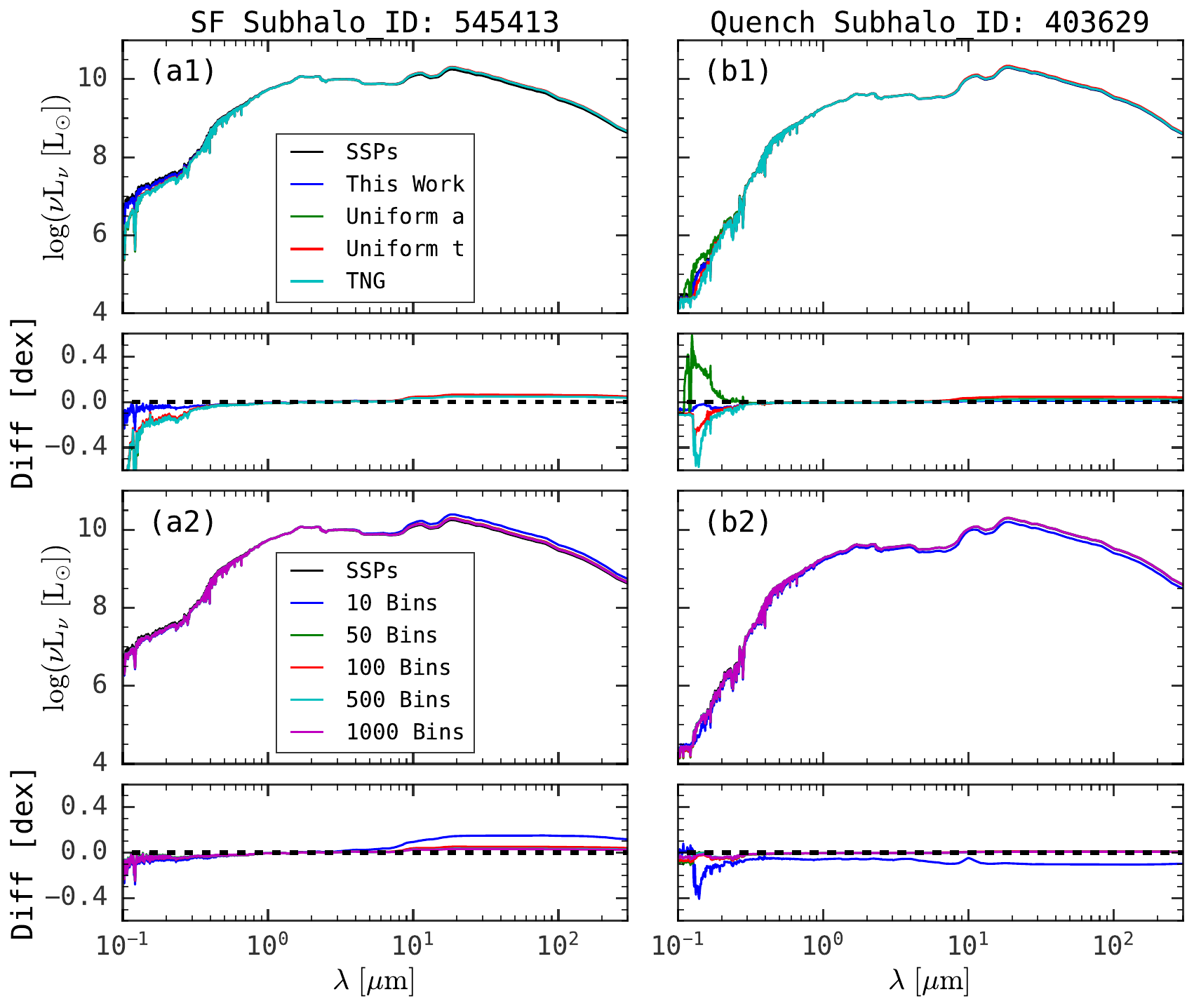}
    \caption{Spectra generated with \textsc{fsps} using various SFH binning algorithms and bin numbers are compared. Both star-forming and quenched galaxies are included, with Subhalo IDs 545413 and 403629, from left to right, respectively.
        The spectra discrepancies between different methods and real spectra generated by combining stellar particles as single stellar populations are quantified in dex in the lower panels. 
        The top two figures (a1, b1) present the comparison between different SFH binning algorithms with a fixed number of 100 bins.
        The blue lines correspond to spectra with the SFH binning algorithm employed in this study. The green and red lines represent spectra generated using uniform scale factors and a uniform time SFH binning algorithm, respectively. The cyan lines demonstrate the binning method that utilizes TNG snapshots as bin edges.
        The bottom figures (a2, b2) display the spectra with varying bin numbers using our method, ranging from 10 bins to 1000 bins.
        This figure illustrates that our method, utilizing 100 bins, effectively recovers UV-to-IR spectra with minimal bias (less than 0.1 dex).}%
    \label{fig:sfh_bin}
\end{figure*}

As shown in Figure~\ref{fig:sfh_bin}, we evaluate the SFH binning algorithms by comparing their ability to reconstruct the real spectra of a star-forming galaxy (Subhalo $\rm ID = 545413$) and a quenched galaxy (Subhalo $\rm ID = 403629$). The effectiveness of each method is measured by the accuracy with which it captures the spectral details

We first treat each stellar particle in the galaxy as a single stellar population and combine them to obtain the most accurate real spectra. We then evaluate different SFH binning methods in the upper panels of Figures~\ref{fig:sfh_bin}(a1) and (b1) with 100 bins, including methods used in this study, bins with uniform scale factor $a$, and uniform time intervals. We also utilize TNG snapshots as bin edges (and add another bin edge at scale factor $a = 0$). Galaxies are then expressed as 100 single stellar populations formed in the middle of the bin.

For most galaxies, our SFH binning algorithm can recover the real spectra within \qty{0.1}{dex}, while other binning methods can lead to significant bias to the real spectra in the short wavelength range. This is because our method can provide a much smaller bin size near $z=0$, where recent star formation can contribute significantly to UV spectra.
We also investigate the influence of bin numbers for our method in Figures~\ref{fig:sfh_bin}\,(a2) and (b2), with bin numbers ranging from 10 to 1000. Our analysis confirms that using 100 bins optimally balances computational efficiency and the fidelity of spectral recovery within \qty{0.1}{dex}. The results validate our algorithm's capability to reconstruct detailed and accurate SFHs for galaxies, supporting its application in broader astrophysical studies.

\section{Validation of SFH-based Cosmic SFR Density}
\label{app:sfrd_test}

This Appendix assesses the accuracy of deriving cosmic SFR density using the SFHs of galaxies (the fossil record method) from the IllustrisTNG simulation compared to direct snapshot-based methods.

\begin{figure*}
    \centering
    \includegraphics[width=0.9\linewidth]{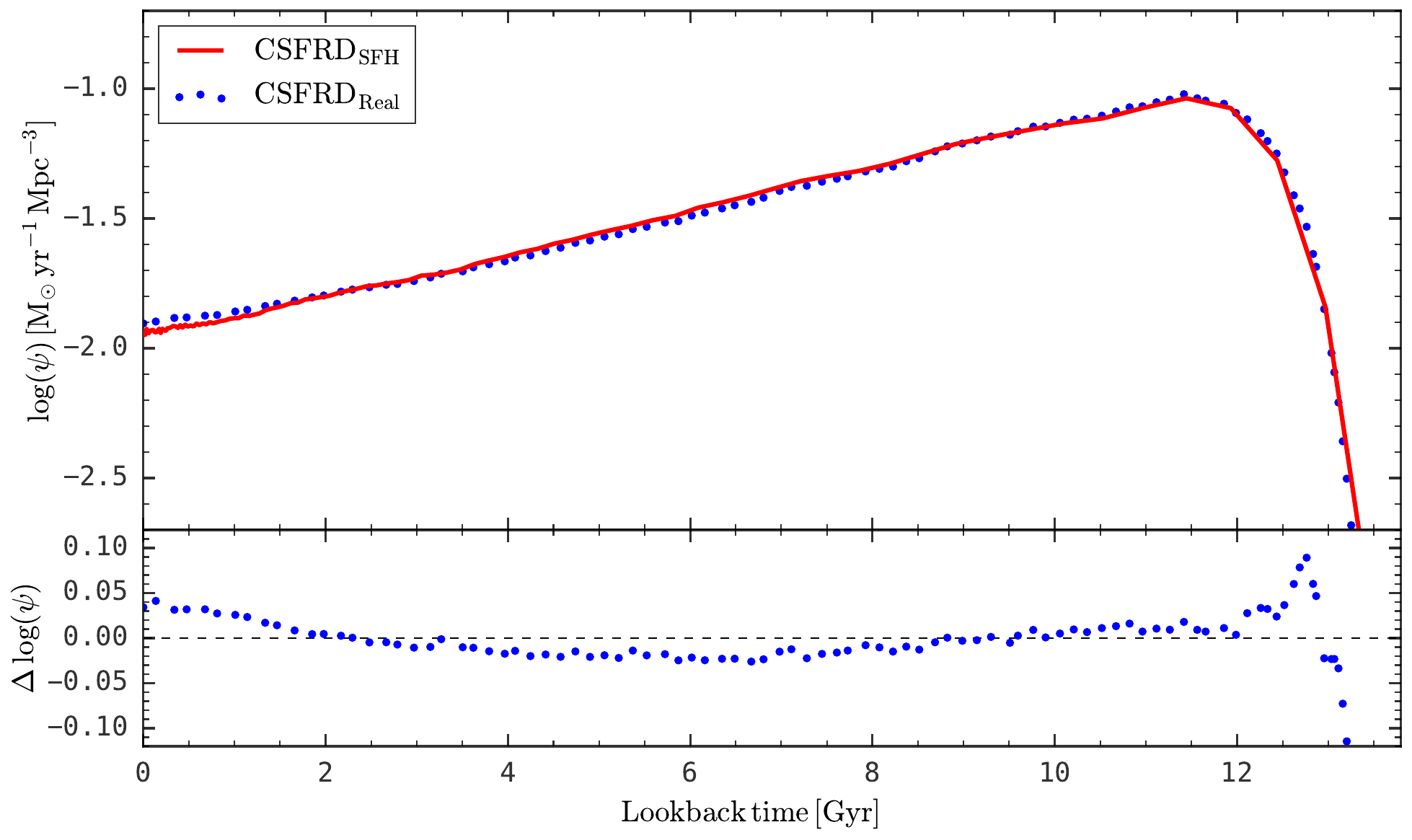}
    \caption{Comparison between SFH-based and direct measurement of cosmic SFR density in TNG100. In the upper panel, the red line shows the cosmic SFR density calculated as the sum of the SFHs of all TNG100 galaxies with a stellar mass more than $10^9\, h^{-1}\, M_{\odot}$ at $z=0$, while the blue dots depict the direct measurements of SFRs from galaxies in each TNG100 snapshot. Differences are highlighted in the lower panel, showing minor discrepancies (less than \qty{0.05}{dex}).}%
    \label{fig:tng_sfrd}
\end{figure*}

Cosmic SFR density is traditionally estimated from IR and UV luminosity functions representing direct observations across various epochs. The SFH-based approach applied in this work, aggregating historical star formation data from simulations, provides an alternative estimation method. The latter approach gives a smaller value than the former because of the contribution of intracluster light and small galaxies. Here, we quantify the difference between these two approaches using TNG100 in Figure~\ref{fig:tng_sfrd}.

The upper panel shows the SFH-based SFR density (red line) against the directly measured ones from TNG100 snapshots (blue dots). The lower panel illustrates their differences, predominantly less than \qty{0.05}{dex}, suggesting that our SFH-based method captures the cosmic SFR density with minimal bias, despite minor deviations in early epochs due to coarse temporal resolution in SFH sampling. This supports the robustness of using integrated SFH for cosmic SFR density estimation in cosmological studies.

\section{Evaluating Dust Attenuation on Spectral Indices}
\label{app:spec_idx_test}
In this Appendix, we investigate the effects of dust attenuation on spectral indices, which were not considered in our initial stellar population synthesis. We assess how dust impacts the indices $D_{\rm n}(4000)$ and $\rm H\delta_A$, which are critical for interpreting the age and activity of stellar populations.

\begin{figure*}
    \centering
    \includegraphics[width=0.7\linewidth]{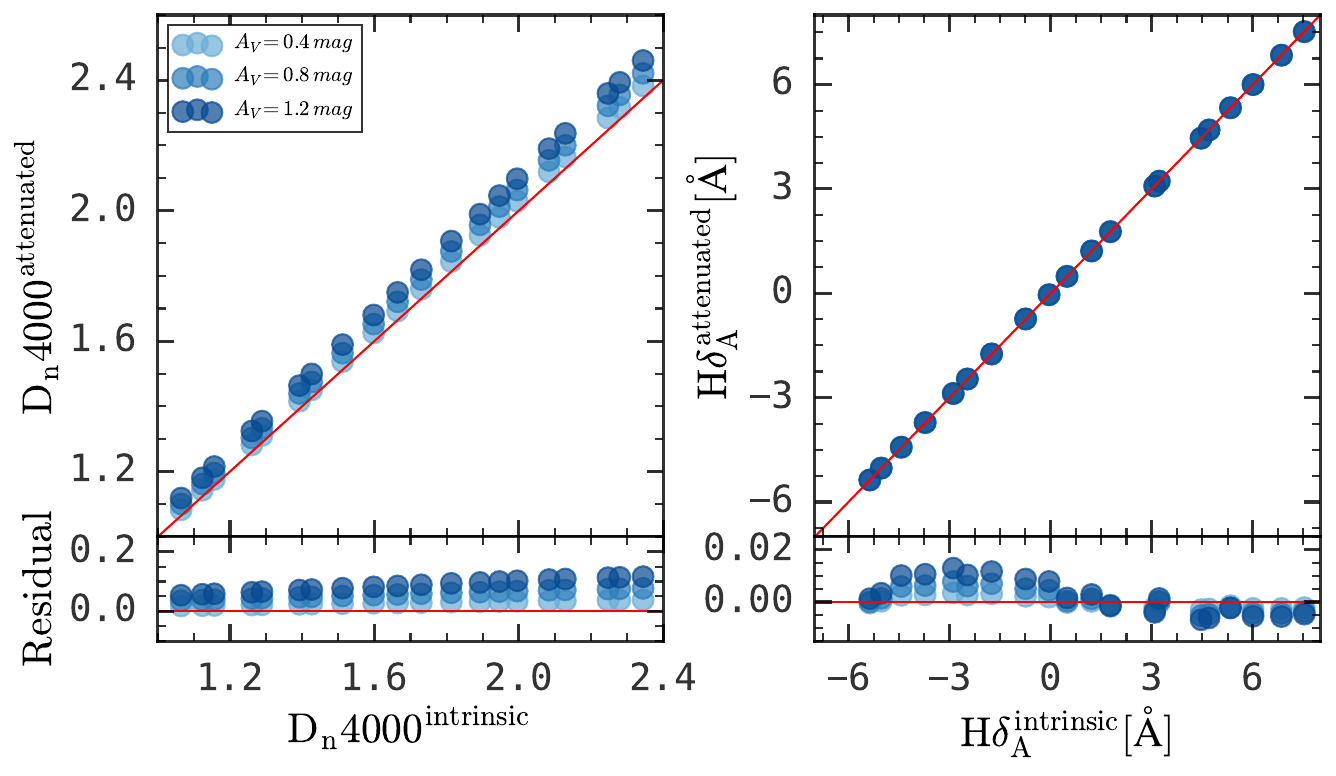}
    \caption{Impact of dust attenuation on spectral indices $D_{\rm n}(4000)$ and $\rm H\delta_A$:
        The upper panels compare intrinsic and dust-affected indices across 20 randomly selected TNG100 galaxies. Galaxies with different $A_V$ are indicated by different colors. The red solid lines are the one-to-one reference lines. The lower panels quantify the differences. These two spectral indices are insensitive to dust attenuation, with the largest changes being $\lesssim 0.15$ for $D_{\rm n}(4000)$ and $\lesssim 0.02$\AA\ for $\rm H\delta_A$.}%
    \label{fig:d4n_hda_av}
\end{figure*}

Using \textsc{fsps}, we generate mock spectra for a representative sample of TNG100 galaxies and apply the extinction law from \citet{Calzetti_2000} to simulate dust effects. This process involves varying levels of attenuation ($A_V = 0.4, 0.8$, \qty{1.2}{mag}) to cover a broad range of realistic conditions \citep[See Figure 8(b) in][]{Salim_2020}. Our findings in Figure~\ref{fig:d4n_hda_av} confirm that both $D_{\rm n}(4000)$ and $\rm H\delta_A$ are relatively robust against dust, with only minor deviations observed even under a significant dust presence. For $\rm D_n4000$, attenuation raises the value, and this effect becomes stronger for higher intrinsic $\rm D_n4000$ galaxies. For $\rm H\delta_A$, there seems to be no clear trend. This reinforces the utility of these indices in dusty environments and aligns with findings from prior studies, such as those by \citet{Kauffmann_2003}, highlighting their reliability in diverse galactic conditions.

\end{document}